\documentclass[runningheads]{llncs}

\usepackage[T1]{fontenc}
\usepackage[utf8]{inputenc}  % Only if you're not using LuaLaTeX/XeLaTeX
\usepackage{times}           % Springer recommends Times

\usepackage{amsmath, amssymb}

\usepackage{graphicx}
\usepackage{subcaption}
\usepackage{booktabs}
\usepackage{multirow}
\usepackage{caption}
\usepackage{enumitem}

\usepackage{xcolor}

\usepackage{longtable}
\usepackage{microtype}

\usepackage[hidelinks,breaklinks=true]{hyperref}
\title{Data-Driven Approach to Capitation Reform in Rwanda}

\author{
Babaniyi Olaniyi\inst{1} \and
Ina Kalisa\inst{2} \and
Ana Fernández del Río\inst{1} \and
Jean Marie Vianney Hakizayezu\inst{1} \and
Enric Jané\inst{1} \and
Eniola Olaleye\inst{1} \and
Juan Francisco Garamendi\inst{1} \and
Ivan Nazarov\inst{1} \and
Aditya Rastogi\inst{1} \and
Mateo Diaz-Quiroz\inst{1} \and
África Periáñez\inst{1}\thanks{africa@causalfoundry.ai (corresponding author)} \and
Regis Hitimana\inst{2}\thanks{regis.hitimana@rssb.rw (corresponding author)}
}

\institute{
Causal Foundry, Barcelona, Spain \and
Rwanda Social Security Board, Kigali, Rwanda
}

\authorrunning{Olaniyi et al.}
\titlerunning{Data-Driven Approach to Capitation Reform in Rwanda}

\begin{document}
\maketitle
%\updateddate

%TODO: add references on capitation, health system reform, antibiotic use, fraud detection, methodologies, our papers, papers Emily recommended, etc
% ======================================================================

\begin{abstract}
As part of Rwanda’s transition toward universal health coverage, the national Community-Based Health Insurance (CBHI) scheme is moving from retrospective fee-for-service reimbursements to prospective capitation payments for public primary healthcare providers. This work outlines a data-driven approach to designing, calibrating, and monitoring the capitation model using individual-level claims data from the Intelligent Health Benefits System (IHBS). We introduce a transparent, interpretable formula for allocating payments to Health Centers and their affiliated Health Posts. The formula is based on catchment population, service utilization patterns, and patient inflows, with parameters estimated via regression models calibrated on national claims data. Repeated validation exercises show the payment scheme closely aligns with historical spending while promoting fairness and adaptability across diverse facilities. In addition to payment design, the same dataset enables actionable behavioral insights. We highlight the use case of monitoring antibiotic prescribing patterns, particularly in pediatric care, to flag potential overuse and guideline deviations. Together, these capabilities lay the groundwork for a learning health financing system: one that connects digital infrastructure, resource allocation, and service quality to support continuous improvement and evidence-informed policy reform.

%We highlight two applied use cases: (1) monitoring antibiotic prescribing patterns, particularly in pediatric care, to flag potential overuse and guideline deviations, and (2) identifying anomalies in pharmacy claims that may indicate fraud. Together, these capabilities lay the groundwork for a learning health financing system: one that connects digital infrastructure, resource allocation, and service quality to support continuous improvement and evidence-informed policy reform.
\end{abstract}

% ======================================================================
\section{Introduction}
\label{sec:introduction}

\begin{figure}[ht]
\centering
\includegraphics[width=1\textwidth]{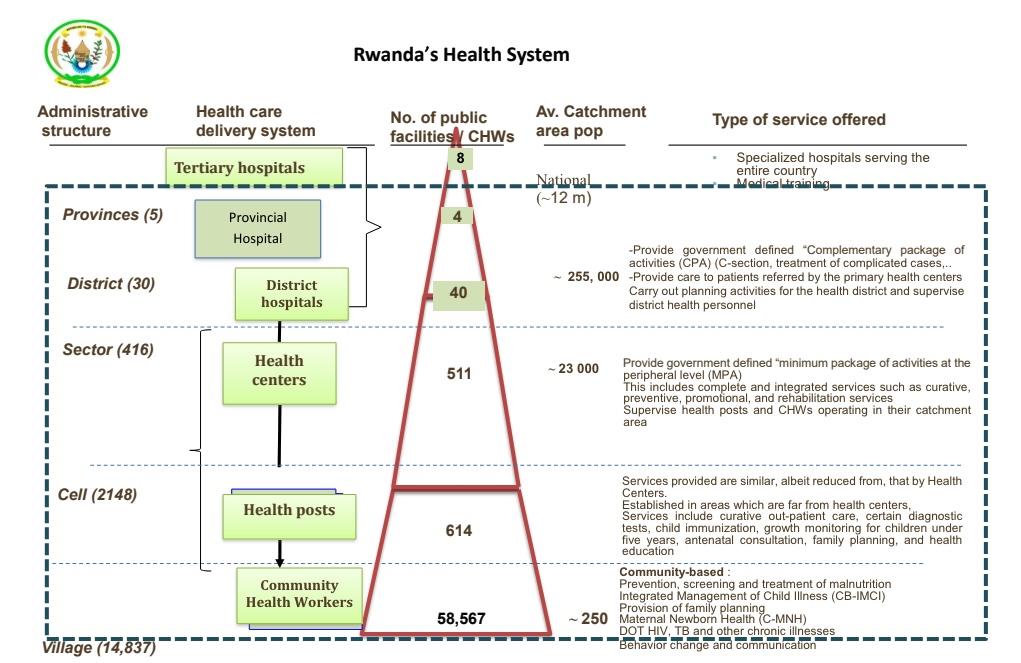}
\caption{\textit{Schematic representation of Rwanda's health system, aligned with its administrative structure. Average catchment populations and number of facilities are indicated at each level.}}
\label{fig:health-system}
\end{figure}

Achieving universal health coverage (UHC) requires not only expanding insurance enrollment but also establishing financing mechanisms that are equitable, efficient, and sustainable. In Rwanda, the national Community-Based Health Insurance (CBHI) scheme covers the vast majority of the population and plays a central role in delivering primary care. Historically, CBHI has reimbursed health facilities using a fee-for-service (FFS) model, where payments are based on the number and type of services delivered. While this approach supports detailed cost tracking, it also creates inefficiencies: administrative burden, delays in reimbursement, and financial incentives that can encourage overtreatment and inflate costs. Under FFS, median cost per visit at PHC level is around 1 USD. However, this cost varies widely: the interquartile range spans 0.70–1.50 USD, with outliers exceeding 3 USD.

To address these challenges, the Rwanda Social Security Board (RSSB), which administers both CBHI and the Rwandaise d’Assurance Maladie (RAMA) scheme for formal sector workers, is transitioning toward a capitation model. Under capitation, facilities receive a fixed prospective payment, enabling more predictable budgeting and shifting the focus toward preventive, cost-effective care. From a facility’s perspective, capitation reduces paperwork and facilitates planning. For the insurer, it helps contain costs and limits exposure to overbilling and fraud. As a result, capitation holds promise as a key enabler of accountability and long-term sustainability in Rwanda’s health financing landscape.

This transition is supported by major investments in digital infrastructure. In 2023, RSSB launched the Intelligent Health Benefits System (IHBS), a national platform for processing PHC claims. Initially deployed for CBHI, IHBS has since been extended to handle some claims for RAMA as well. This enables a capitation model grounded in actual service utilization, responsive to local variations in population need, and adaptable over time. It also supports broader system monitoring through behavioral analytics.

This work documents the implementation of Rwanda’s national capitation model for CBHI-financed primary care. It presents:

\begin{itemize}
   \item A transparent, data-driven payment formula based on IHBS data, tailored to heterogeneous facility catchments and referral patterns, with mechanisms for calibration, validation, and operational monitoring.

    \item One use case illustrating the broader potential of IHBS for behavioral insight: A focus on antibiotic prescription patterns in pediatric care.
\end{itemize}

Together, these components represent a foundation for a learning health financing system: one in which digital infrastructure, funding flows, and service delivery are connected through data and continuous policy feedback.

\begin{figure}[ht]
\centering
% \includegraphics[width=0.8\textwidth]{Figures/map_hc_closest_hospital.pdf}
% (2025-10-22 171X) vector -> raster using `magick -density 300 map_file.pdf -resize 30% map_file.png`
\includegraphics[width=0.8\textwidth]{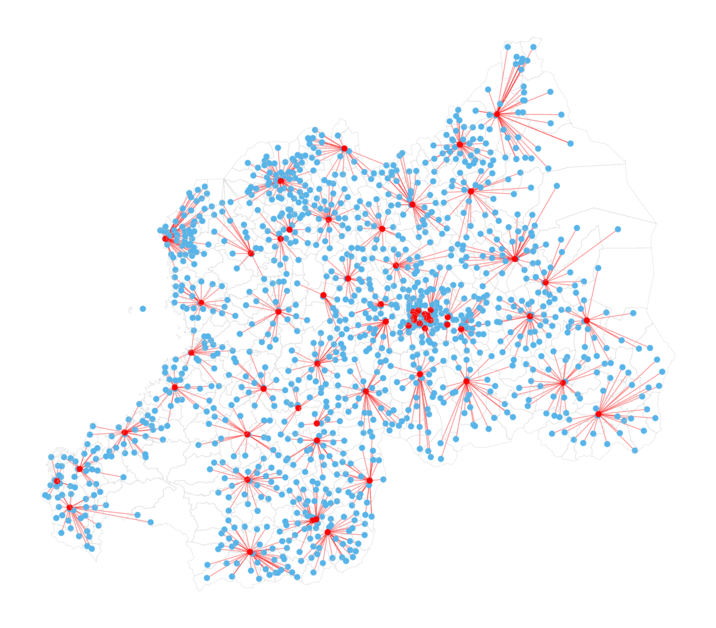}
\caption{\textit{Health Centers connected to the nearest hospital across sectors. The Health Centers (blue dots) and the nearest Hospitals (red dots), along with the line connecting the health centers to the nearest hospital (red lines)}}
\label{fig:hc_closest_hospital}
\end{figure}

\section{Primary Healthcare Delivery and Financing in Rwanda}

Rwanda mandates health insurance for all residents, with CBHI covering the majority. CBHI beneficiaries access a broad primary care package through public Health Centers (HCs) and Health Posts (HPs), supported by modest copayments and tiered annual contributions. Other public insurance schemes, such as the Rwandaise d'Assurance Maladie (RAMA), which serves civil servants and formal sector employees, cover most of the remaining population. Both CBHI and RAMA are administered by the Rwanda Social Security Board (RSSB), positioning it as the central public insurer responsible for health coverage across nearly the entire population.

\begin{figure}[ht]
\centering
\includegraphics[width=0.8\textwidth]{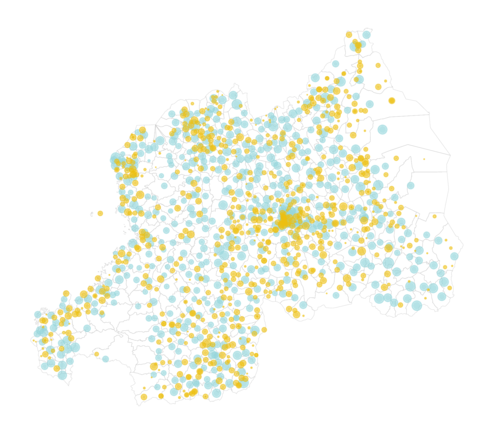}
\caption{\textit{The geographical locations of facilities across their sectors are shown, indicating Health Centers (in blue) and private Health Posts (in yellow). The size of each facility is proportional to the volume of visits}}
\label{fig:health-centers}
\end{figure}

As shown in Figure~\ref{fig:health-system}, Rwanda's healthcare delivery system is tiered and decentralized, mirroring the country's administrative structure. Health Centers (approximately 500 nationwide) operate at the sector level and supervise affiliated Health Posts (roughly 1,700, public and private combined), delivering the Minimum Package of Activities (MPA). Community Health Workers (CHWs), numbering over 50,000, form the frontline layer. District and provincial hospitals provide more complex care. Together, HCs and HPs resolve over 90\% of care episodes for Rwanda's 14.1 million citizens \cite{NISR}. Figure \ref{fig:hc_closest_hospital} shows a map with the location of health centers connected to the location of their nearest hospital, while HC and HP location and relative visit volume are shown in Figure~\ref{fig:health-centers}.

Health Centers may manage one or more affiliated Health Posts, which function as operational extensions of the parent facility. These managed HPs typically deliver a limited set of basic services and fall under the administrative and reporting structure of the supervising Health Center. However, service delivery data do not distinguish between visits conducted at HCs versus those conducted at their affiliated HPs, and the number of HPs managed by each HC is not systematically recorded. 

In addition to these public Health Posts, there are privately operated Health Posts that function independently within the same geographic catchments. These facilities serve the same population but are not managed by the Health Center and generally differ in service offerings, operational procedures, and incentives. The scope of the capitation model comprises the Health Center and its managed Health Post extensions, excluding for now private HPs. However, visits to both public and private Health Posts are included when calculating catchment-level utilization metrics.

Historically, CBHI reimbursed PHC facilities via monthly, manually verified FFS claims. This model introduced delays in reimbursement to facilities, constrained scalability (e.g., two RSSB staff members were required per facility for claims processing), and created opportunities for inefficiency. These challenges have prompted the national insurer to transition to capitation for PHC, with rollout beginning in November 2025 in the Eastern Province and then gradually expanding to the remaining provinces. The capitation model applies to each Health Center and its affiliated public Health Posts, using historical utilization and catchment data to estimate prospective payments.

Under capitation, each Health Center will be responsible for delivering a standardized package of primary health care services to the population within its designated catchment area. These catchments are defined at the village level and assigned by the Ministry of Health (MoH), although CBHI beneficiaries may also seek care outside their assigned catchment. Some Health Centers are designated as Medicalized Health Centers because they provide additional services, such as cesarean sections, that fall outside the standard PHC package. These services are reimbursed separately through FFS and are excluded from the capitation model described in this article.

\begin{figure}[ht]
\centering
\includegraphics[width=0.9\textwidth]{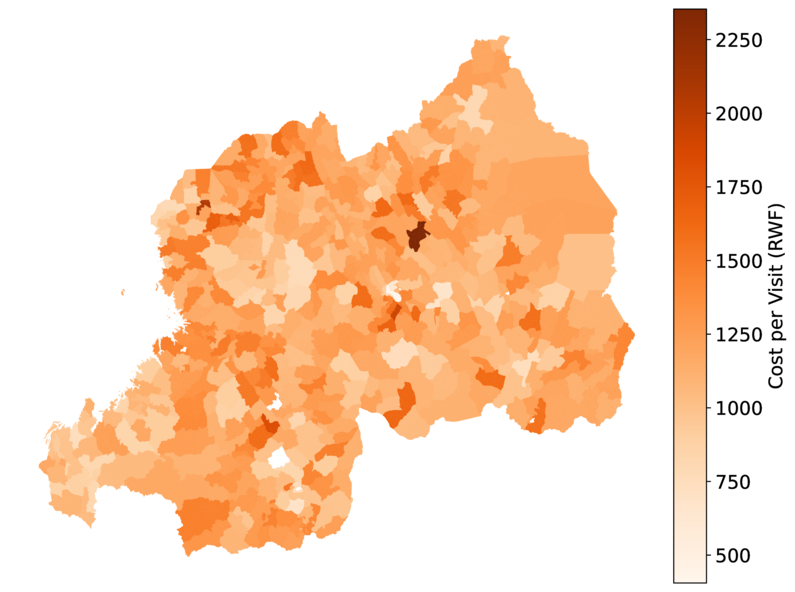}
\caption{\textit{Average cost per visit across non-medicalized Health Centers.}}
\label{fig:visit-cost}
\end{figure}

% ======================================================================
\section{The Capitation Reform}
\label{sec:capitation}

Data-driven capitation design is enabled by the rich data captured through IHBS, which allows for capitation payments  grounded in real-world utilization patterns and facility-level monitoring, dynamic adjustment, and policy feedback loops. All analyses documented in this work were performed on anonymized datasets in compliance with national data protection regulations, preserving confidentiality while ensuring modeling fidelity.

Since April 2023, this in-house digital platform is used to manage all CBHI claims at the primary care level. IHBS supports the full lifecycle of reimbursement, from consultation recording to monthly invoice submission, review, and approval, and captures rich, structured data at the visit level. For each encounter, the system logs patient demographics, diagnoses, procedures, prescribed items, costs, and the identity of healthcare personnel involved. In addition to clinical and financial data, IHBS integrates administrative information, including household location (sourced from the Ministry of Local Government) and catchment area assignments (as defined by the MoH). It also links to Rwanda's new Imibereho social registry, which classifies households into socioeconomic categories based on predefined indicators. Since February 2025, IHBS is also used by pharmacies to submit their claims on drugs dispensed to RAMA members

The capitation formula was designed to reflect both intra-catchment and inter-catchment patient flows, ensure equitable funding across heterogeneous settings, and differentiate between standard and medicalized Health Centers. It supports dynamic recalibration based on updated utilization data, enabling RSSB to adjust payments in response to seasonal or structural changes. The resulting framework offers a transparent and scalable basis for PHC financing aligned with Rwanda's universal health coverage goals.

To ensure practical adoption and institutional alignment, the formula was developed according to three core principles:

\begin{enumerate}
    \item \emph{Simplicity and transparency.} The model avoids opaque scoring systems or complex statistical constructs. Parameters are interpretable and grounded in measurable system attributes, facilitating understanding by facility managers.
    
    \item \emph{Behavioral neutrality.} The formula excludes variables that could incentivize undesirable practices. For example, inpatient admissions were deliberately omitted to avoid encouraging unnecessary hospitalizations. Predictors are limited to factors largely outside the control of individual facilities, such as catchment population size and historical utilization trends.
    
    \item \emph{Budget alignment.} Predicted payments were calibrated to approximate historical expenditures under the FFS model. This design minimizes funding disruptions during the transition, giving facilities time to adapt while preserving continuity of care.
\end{enumerate}

%TODO: update end date of dataset if we update the analyses
These principles shaped a methodology rooted in empirical data, transparency, and operational realism, ensuring that the resulting capitation model aligns financial incentives with health system goals while maintaining provider engagement and service quality. Methodological development combined data engineering, descriptive analytics, and parameter estimation. The dataset used for the analyses presented in this work spanned April 1, 2023 to March 4, 2025, and included visit-level and membership data from IHBS, enriched with household location and catchment definitions.

\begin{figure}[ht]
\centering
\includegraphics[width=0.8\textwidth]{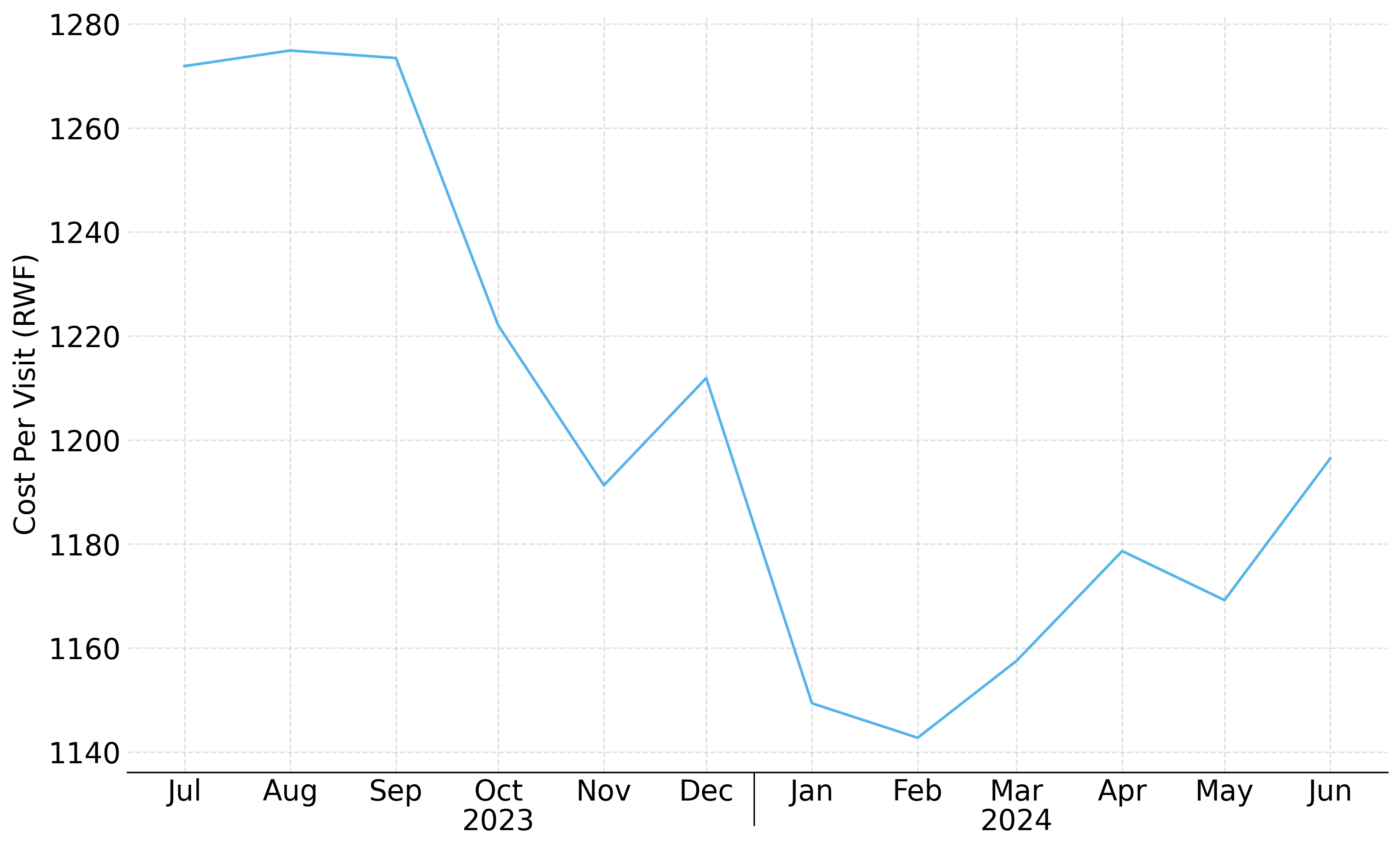}\\
\includegraphics[width=0.8\textwidth]{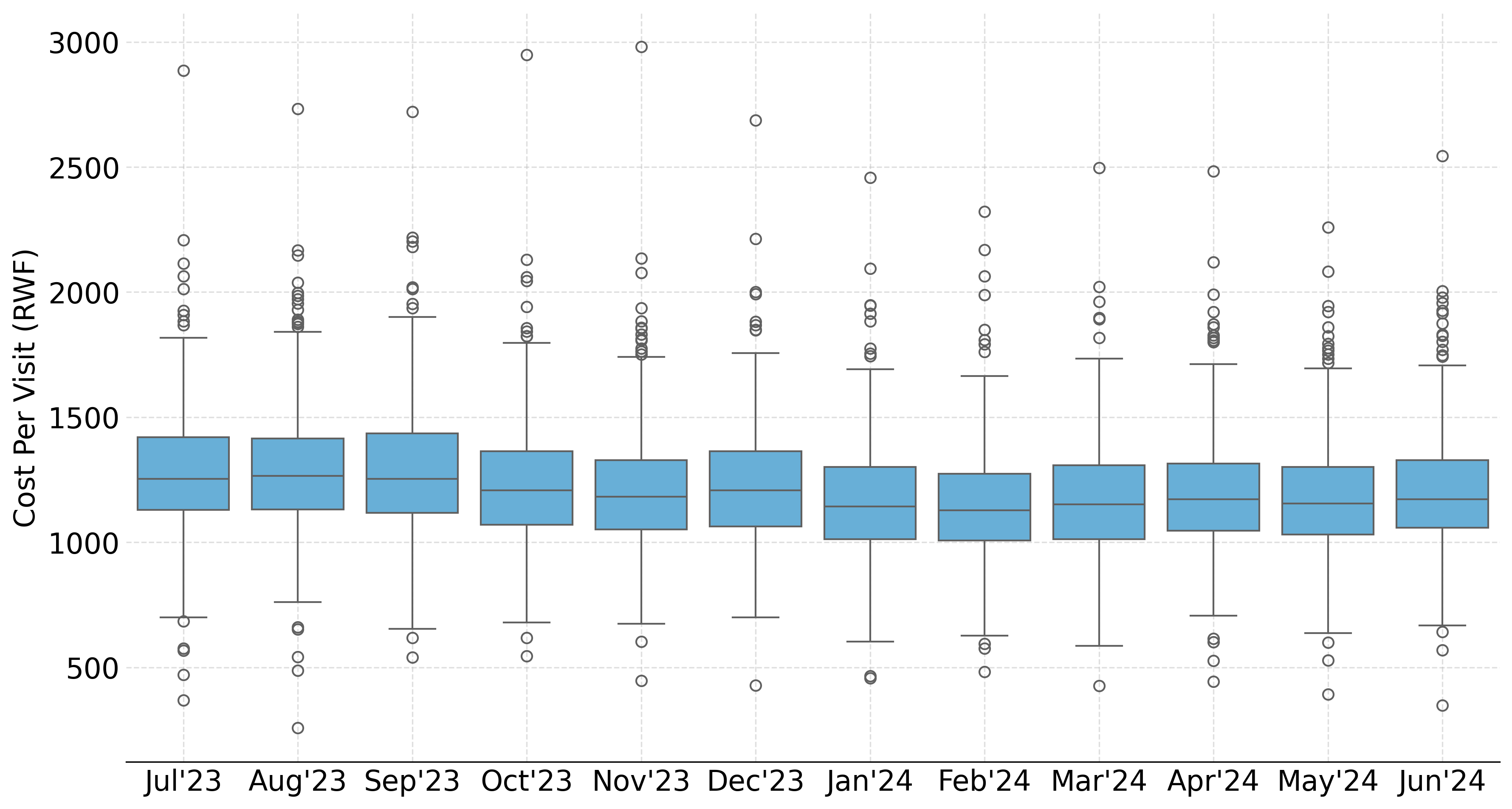}
\caption{\textit{Average monthly cost per visit. Top plot represents the average over all HCs while the bottom plot shows its distribution across facilities.}}
\label{fig:cost-per-visit}
\end{figure}

\subsection{Capitation metrics}
%TODO: add map figures showing PHC utilization, catchment population, health center capture ratio, and inflow

The metrics used to estimate the capitation corresponding to each facility include the total cost associated with visits during a year, the total number of members in the facility's catchment area, the healthcare utilization patterns of members in the catchment area, and the inflow of patients from different catchment areas. Table~\ref{tab:statistics} shows average values for these and other metrics to give an overview of the healthcare capacity and utilization.

%\begin{figure}[ht]
%\centering
% % \includegraphics[width=0.8\textwidth]{Figures/map_inflow.pdf}
%\includegraphics[width=0.8\textwidth]{Figures/raster/map_inflow.png}
%\caption{\textit{Inflow}}
%\label{fig:inflow}
%\end{figure}

%\begin{figure}[ht]
%\centering
% % \includegraphics[width=0.8\textwidth]{Figures/map_hc_capture.pdf}
%\includegraphics[width=0.8\textwidth]{Figures/raster/map_hc_capture.png}
%\caption{\textit{Health Center Capture Ratio}}
%\label{fig:hc_capture}
%\end{figure}

%\begin{figure}[ht]
%\centering
% % \includegraphics[width=0.8\textwidth]{Figures/map_members_catchment.pdf}
%\includegraphics[width=0.8\textwidth]{Figures/raster/map_members_catchment.png}
%\caption{\textit{Members in Catchment}}
%\label{fig:members_catchment}
%\end{figure}

%TODO: Check teh resoultion is sector
\begin{figure}[ht]
\centering

\includegraphics[width=0.45\textwidth]{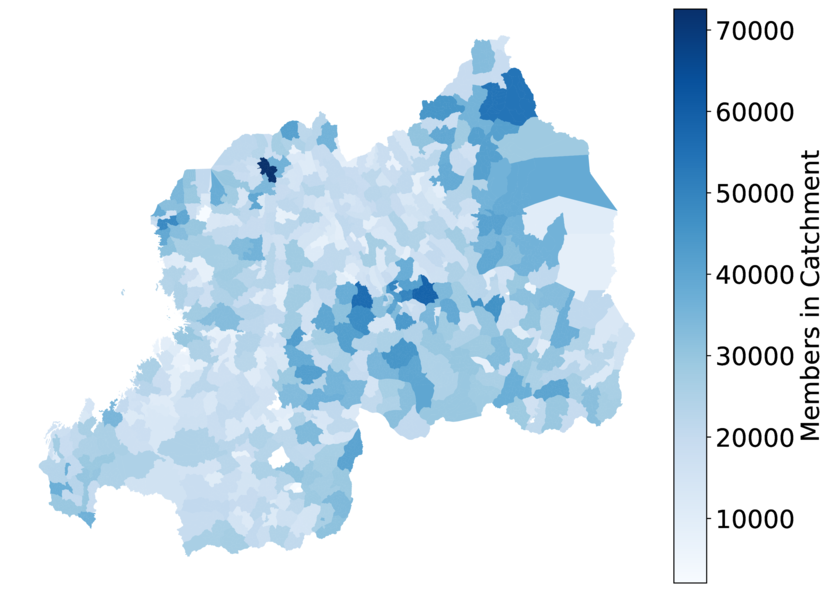}
\includegraphics[width=0.45\textwidth]{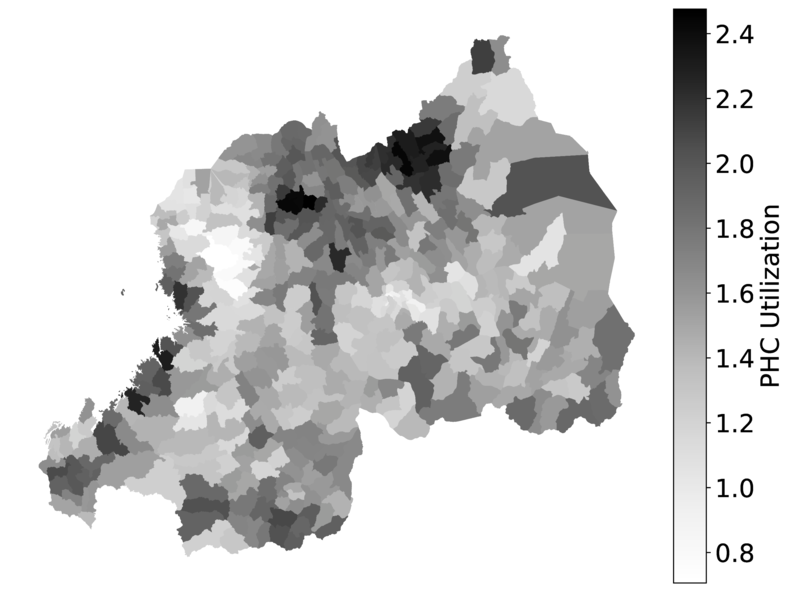} \\
\includegraphics[width=0.45\textwidth]{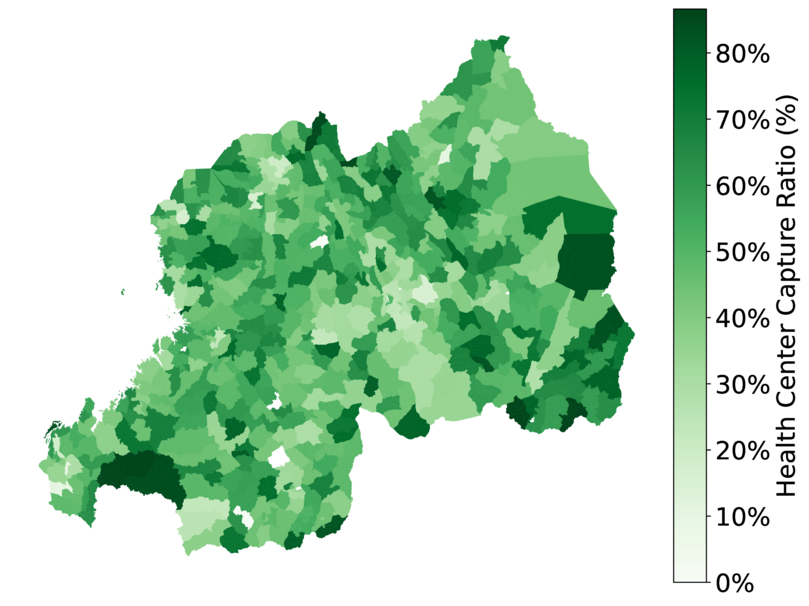}
\includegraphics[width=0.45\textwidth]{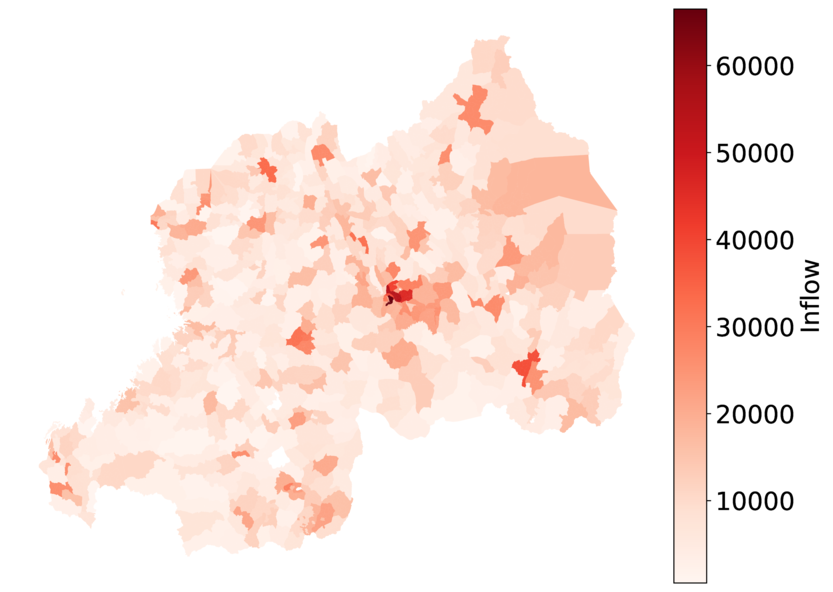}

\caption{\textit{Geographical variation of the metrics used to calibrate capitation payments to past costs across sectors: CBHI members population (top left), PHC utilization rate (top right), health center capture ration (bottom left) and inflow (bottom right)}}
\label{fig:capitation-metrics-map}
\end{figure}

\subsubsection{Total Visit Costs.}

The total annual cost of care per Health Center was calculated by summing the value of services and prescriptions provided to patients, excluding ambulance-related expenses (which will be reimbursed separately under the capitation model). Cost data were extracted from a table listing all cost items associated with each consultation, and the patient co-payment for each visit was deducted. Ambulance co-payments were estimated at 10\% of the ambulance cost and subtracted from the total patient co-payment. To account for instances where the recorded ambulance co-payment was lower than the estimated 10\% (which were considered data input errors), we applied a logic that selected the minimum between the reported co-payment and the estimated co-payment for each visit. Only visits approved by RSSB were included in the computation of the metrics. The annual costs for each facility were adjusted by annualizing the data based on the months of activity, ensuring comparability across Health Centers with partial-year data. Figure \ref{fig:visit-cost} shows a map with the average visit cost across non-medicalized Health Centers, while Figure \ref{fig:cost-per-visit} represents the average monthly visit cost across all facilities (top) and its distribution across facilities (bottom).

Notably, visits that included services outside the standard PHC basket, such as those involving cesarean sections or other procedures typically performed at Medicalized Health Centers, were excluded from the capitation cost calculation. These medicalized services will continue to be reimbursed through the FFS model after the launch of capitation. This ensures that the capitation formula covers only standard PHC services and that facilities offering higher-complexity care are compensated through mechanisms better aligned with the nature of those services. 

\begin{figure}[ht]
\centering
\includegraphics[width=0.8\textwidth]{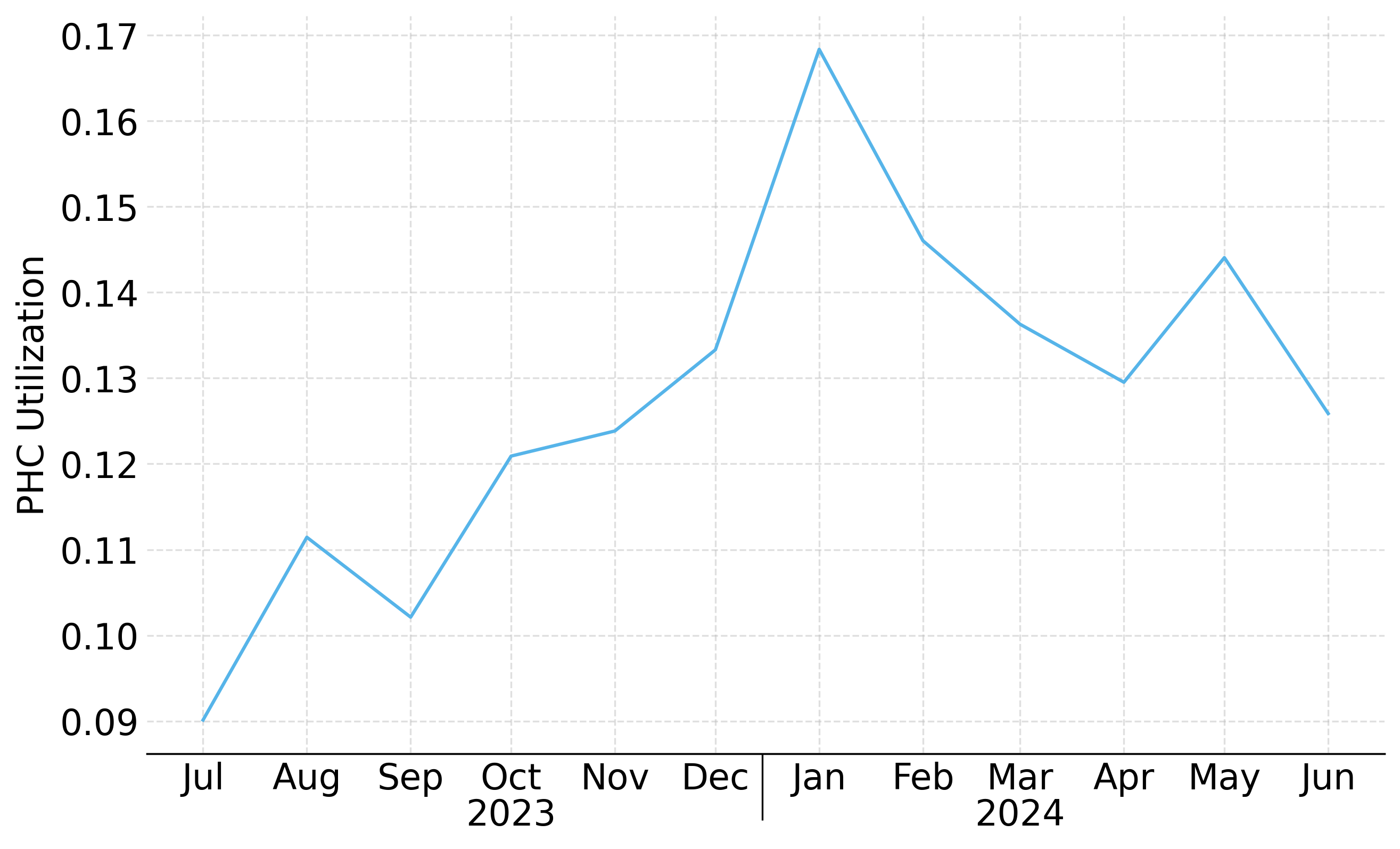}\\
\includegraphics[width=0.8\textwidth]{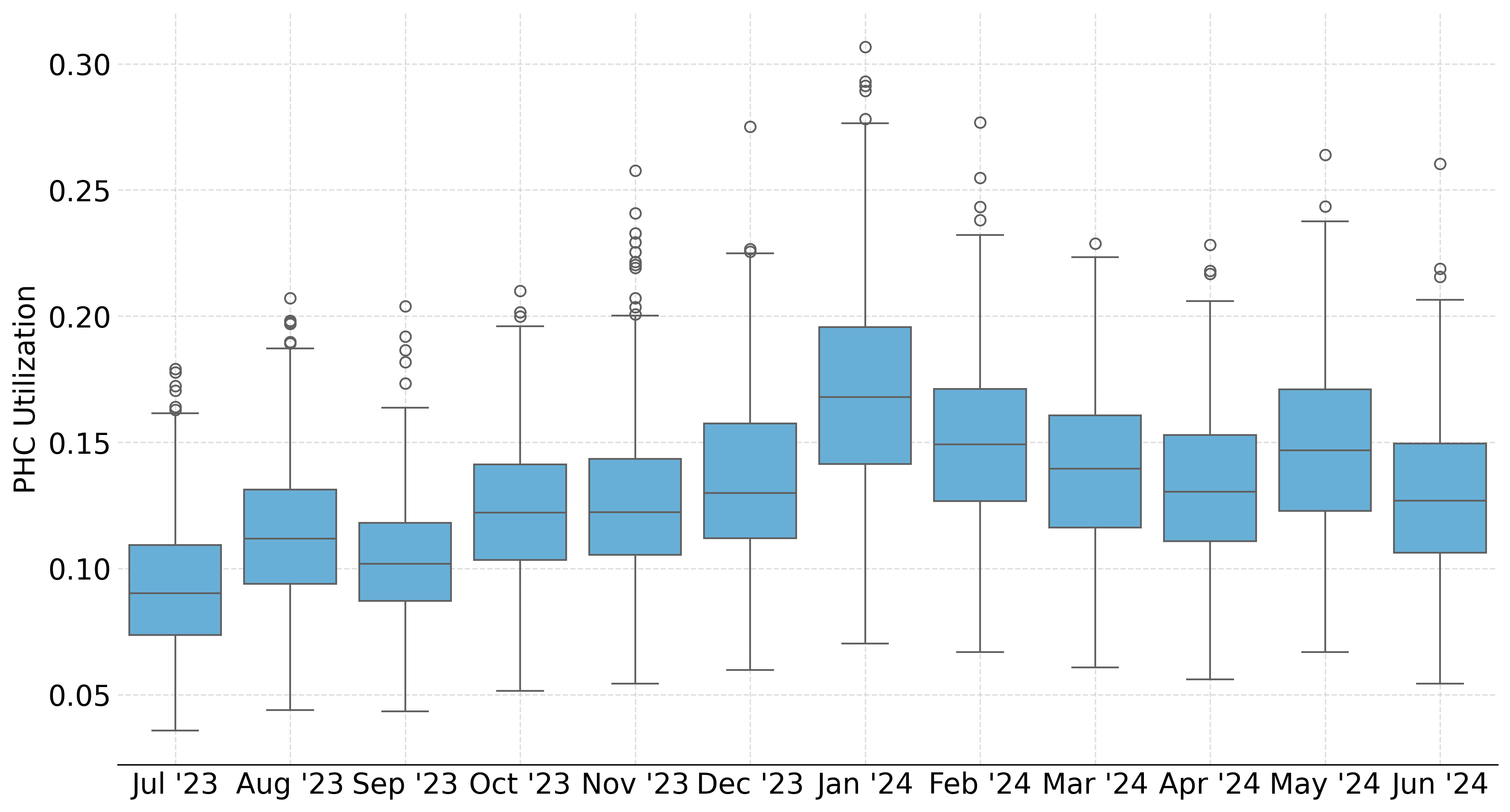}
\caption{\textit{Average monthly PHC utilitation rate. Top plot represents the average over all HCs while the bottom plot shows its distribution across facilities.}}
\label{fig:phc-utilization}
\end{figure}

\subsubsection{Catchment Area Members.}

CBHI membership is at the household level and runs from July through June. Households may pay their contributions at any point during this period, though most do so at the beginning of the cycle. Some members pay only part of the annual fee, which can introduce ambiguity in their active status. Based on discussions with the developers of the IHBS system, any member not marked as inactive but whose record was updated at the start of the semester (January 1 or July 1) is assumed to have paid for the preceding semester and should be considered as active. Therefore, we incorporated this logic when determining the number of CBHI members in each catchment area. Finally, the dataset provided membership information based on the current date; therefore, it was not possible to determine the number of members in each catchment area at any point in time. Thus, the analyses assume that catchment populations remain relatively stable over short periods (e.g., one year) and that errors introduced by variations within short time scales will be minimal. Figure \ref{fig:capitation-metrics-map} top left shows the geographical variation of the number of CBHI members per sector.
%TODO: Check the resoultion is sector

\subsubsection{Healthcare Utilization Metrics.}

The IHBS data were processed to calculate the number of monthly visits by catchment and non-catchment members, disaggregated by location and facility type (Health Centers vs. Health Posts). These visit metrics were used to derive three key indicators for each catchment area:

\begin{figure}[ht]
\centering
\includegraphics[width=0.8\textwidth]{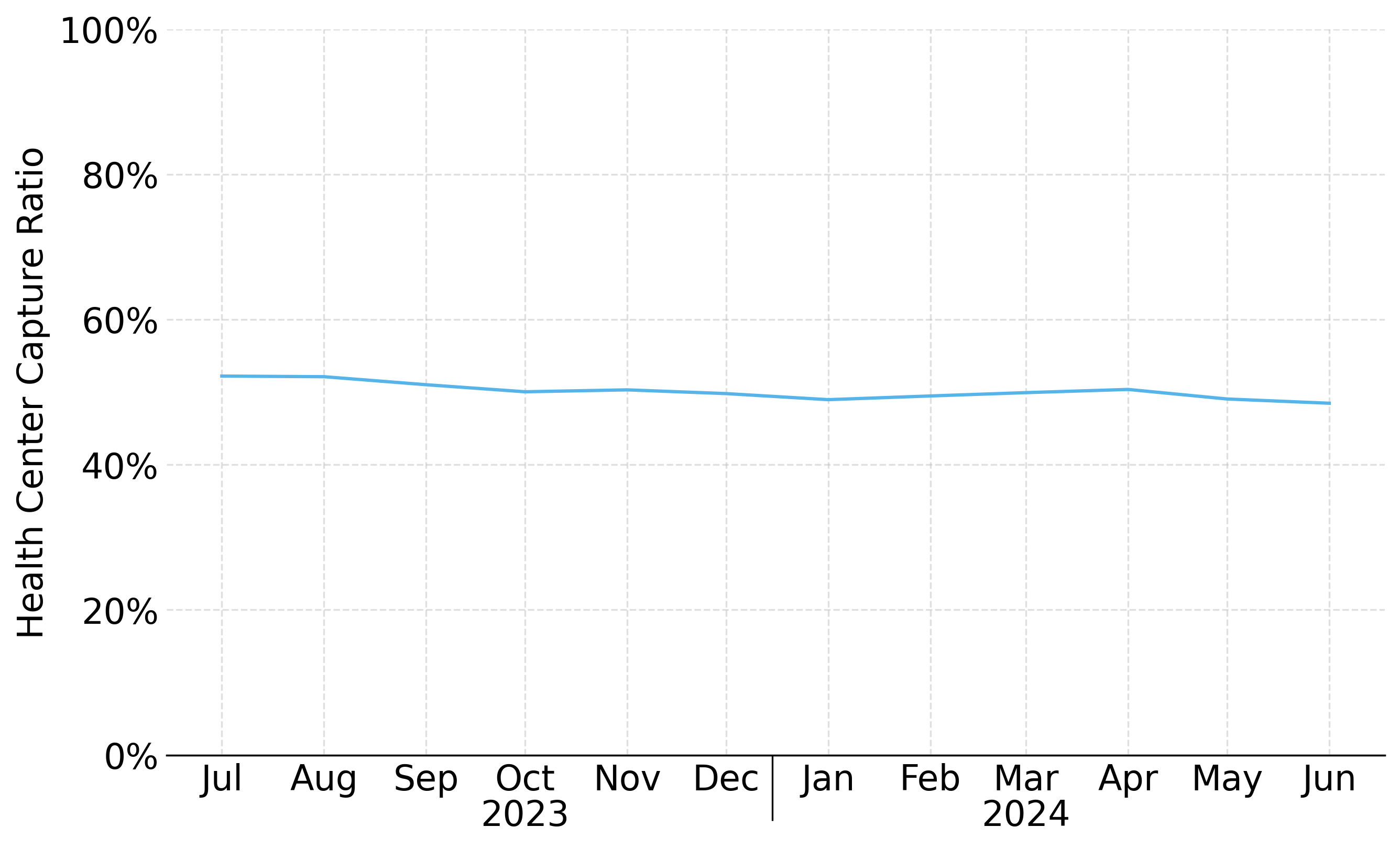}\\
\includegraphics[width=0.8\textwidth]{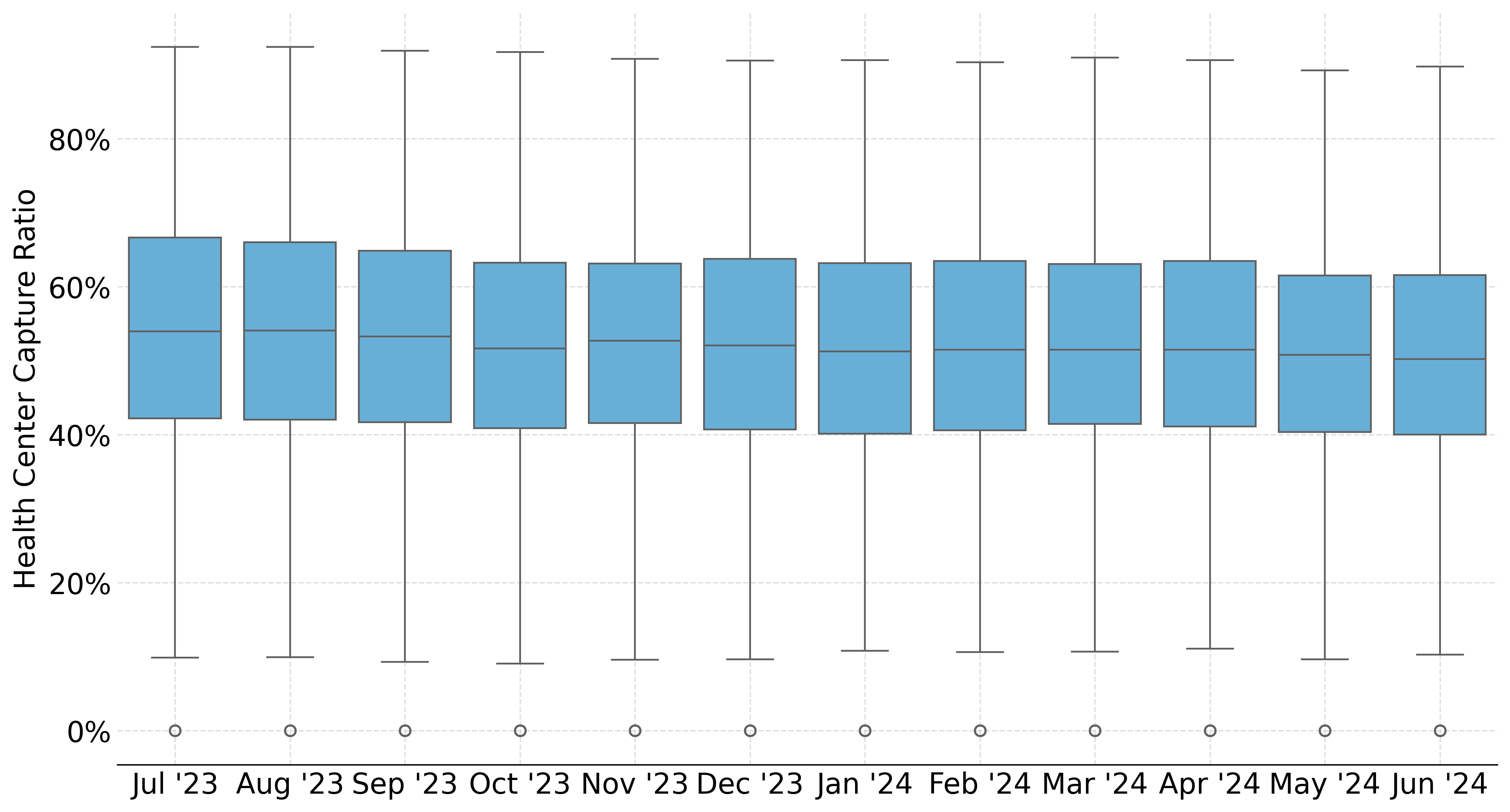}
\caption{\textit{Monthly catchment health center capture ratio. Top plot represents the average over all HCs while the bottom plot shows its distribution across facilities.}}
\label{fig:capture-ratio}
\end{figure}

\begin{figure}[ht]
\centering
\includegraphics[width=0.8\textwidth]{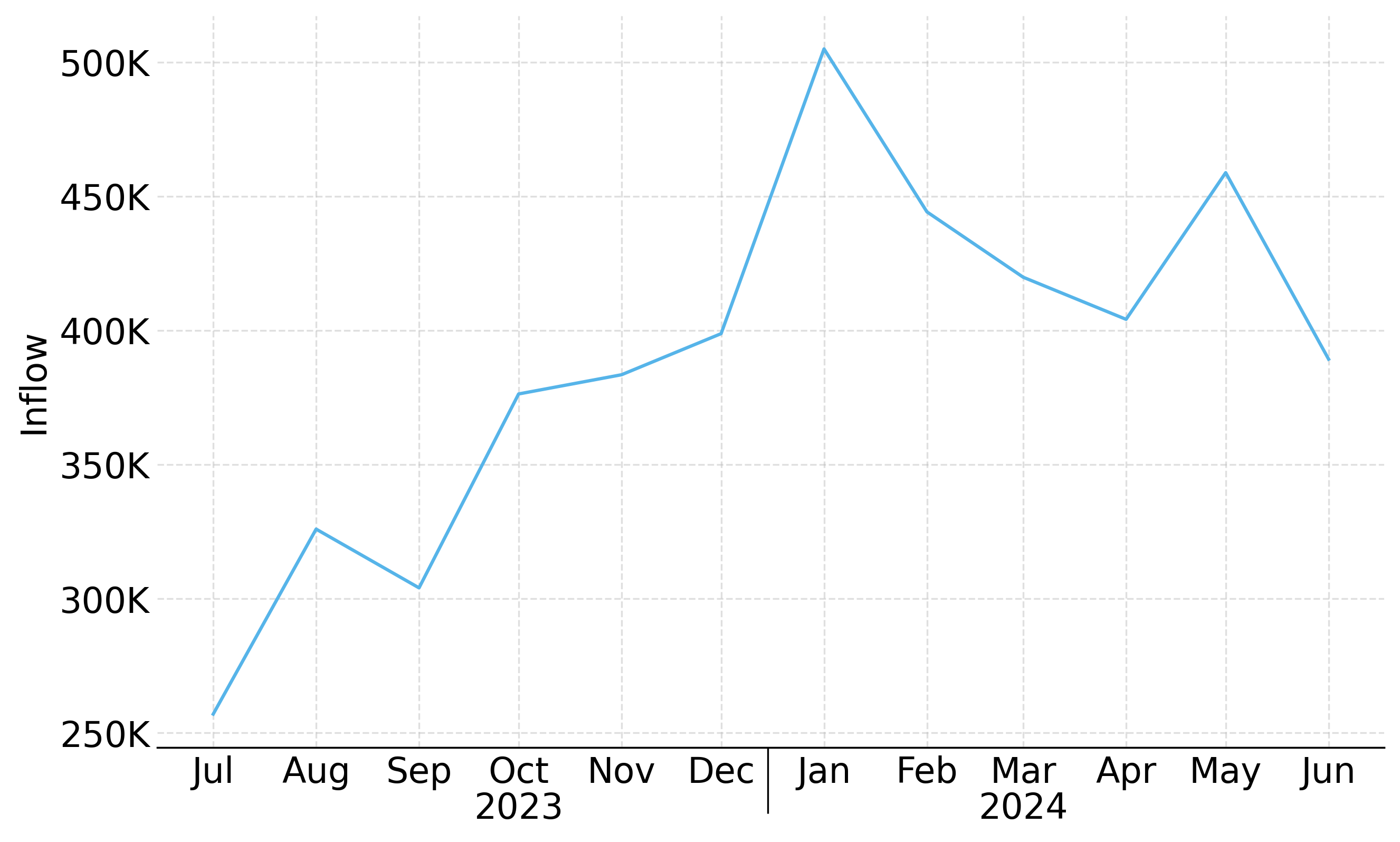}\\
\includegraphics[width=0.8\textwidth]{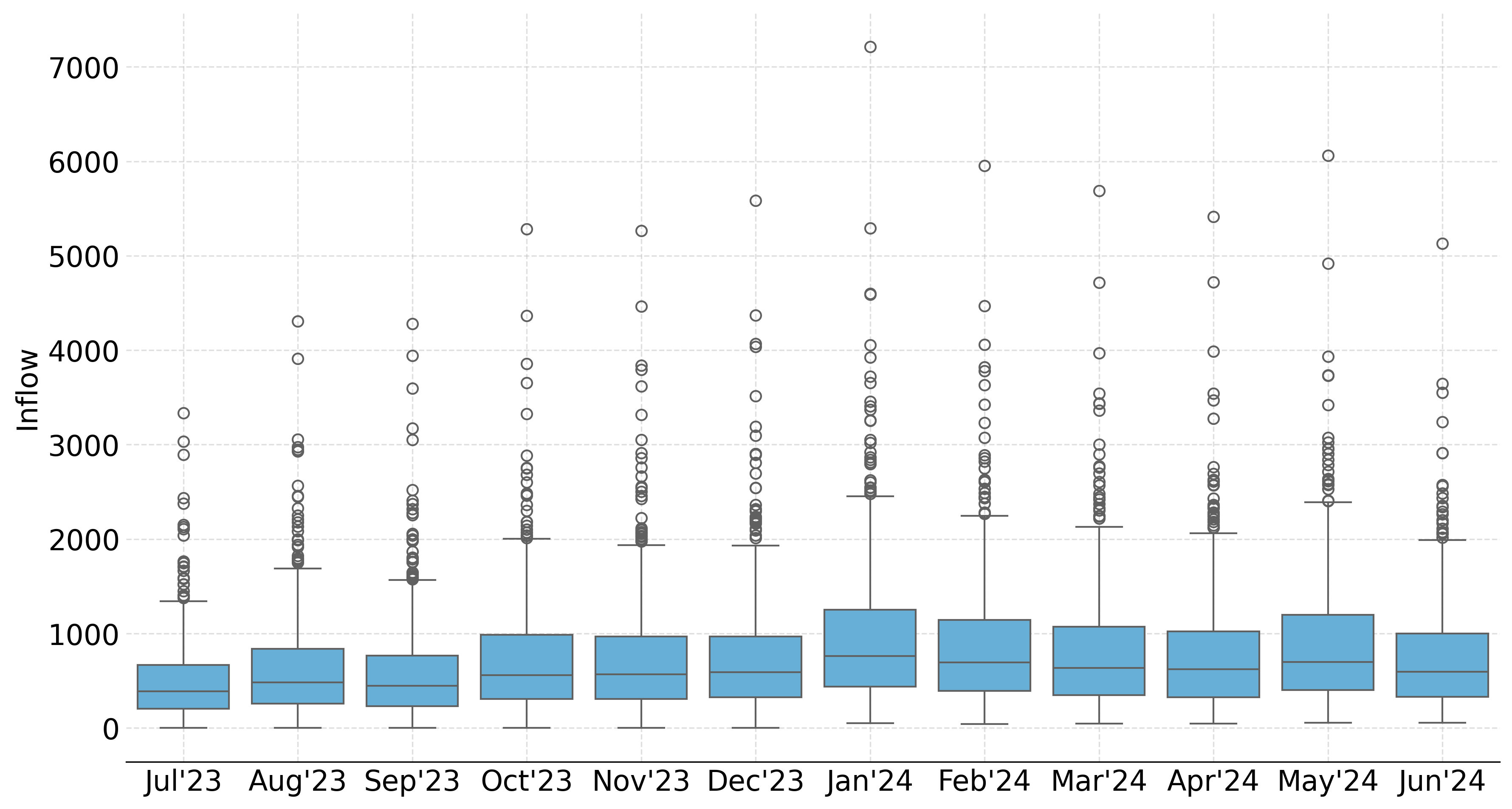}
\caption{\textit{Monthly inflow from members to HC outside their catchment area. Top plot represents the average over all HCs while the bottom plot shows its distribution across facilities.}}
\label{fig:inflow}
\end{figure}

%TODO: Check teh resoultion is sector
\begin{itemize}
    \item \emph{Catchment PHC Utilization Rate:} total annual visits to Health Centers and Health Posts by catchment members, regardless of the location of the facility they visited, divided by the total number of catchment members. The geographical distribution across sectors of this metric is displayed in Figure \ref{fig:capitation-metrics-map} top right. Figure \ref{fig:phc-utilization} shows its monthly average across the country (top) and its monthly distribution across facilities (bottom)).
    \item \emph{Catchment Health Center Utilization Rate:} total annual visits to a Health Center by its catchment members, divided by the total number of catchment members. 
    \item \emph{Health Center Capture Ratio:} proportion of catchment-member visits that occurred at the Health Center in their catchment area. The geographical distribution across sectors of this metric is displayed in Figure \ref{fig:capitation-metrics-map} bottom left. Figure \ref{fig:capture-ratio} shows its monthly average across the country (top) and its monthly distribution across facilities (bottom)).
    \item \emph{Inflow:} total annual visits to the Health Center in the catchment area by patients from other catchment areas. The geographical distribution across sectors of this metric is displayed in Figure \ref{fig:capitation-metrics-map} top right. Figure \ref{fig:inflow} shows its monthly average across the country (top) and its monthly distribution across facilities (bottom)).
\end{itemize}

\begin{figure}[ht]
\centering
\includegraphics[width=0.8\textwidth]{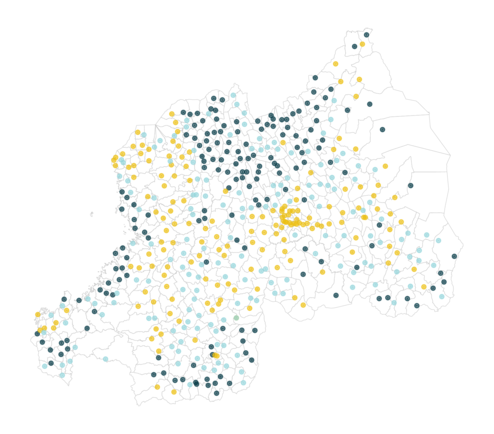}
\caption{\textit{PHC utilization by catchment members, with areas of low utilization colored in yellow, standard utilization in light blue, and high utilization in dark blue}}
\label{fig:phc-utilization-cat}
\end{figure}

Catchment areas were categorized by the \emph{Catchment PHC Utilization Rate} into \texttt{low}, \texttt{medium}, or \texttt{high} utilization, with this category shown for each HC on the map in Figure \ref{fig:phc-utilization-cat}. Furthermore, Health Centers were segmented into five groups based on the \emph{Health Center Capture Ratio}. Within each of those five groups, the median \emph{Catchment Health Center Utilization Rate} was computed. Only visits approved by RSSB were included in the computation of the metrics, and all values were annualized based on the months of activity, ensuring comparability across Health Centers with partial-year data. In both cases, the segmentation is performed ensuring all groups have the same number of facilities.

\subsection{Capitation Equation}
%TODO: why this formula? add plots to justify segmentation and parameters selected, describe the other tests
% (enric) I added some text at the beginning to remind the reader that this was an iterative process starting off from the version RSSB had already developed. There are non-technical reasons to do that, including the fact that building off something developed by RSSB supported ownership and agreement, which was key at that time when we had just started working with them.

%As mentioned previously, 
%TODO: modify data date range if we inclue updated data
The formula builds on an initial proposal developed by RSSB that included two components, one proportional to the members in the catchment, and another component proportional to visits across catchment areas (members visiting Health Centers in other catchment areas, and members from other catchment areas visiting the Health Center of interest). The formula was refined through several iterations, with all stakeholders working closely, by leveraging the IHBS data and exploring other parameters that could potentially help improve the anticipated capitation amount per facility compared to the historical expenditure. %This development combined data engineering, descriptive analytics, and parameter estimation. The dataset spanned April 1, 2023 to March 4, 2025, and included visit-level and membership data from IHBS, enriched with household location and catchment definitions.

The capitation formula includes two components: one that accounts for the number of members in a Health Center's catchment area, adjusted by utilization patterns, and another that compensates for visits from non-catchment patients (\textit{inflow}). For each Health Center, the capitation amount is modeled as:

\[
\text{Capitation} = A_i \cdot U \cdot M + B \cdot I
\]

\noindent where:

\begin{itemize}
  \item $A_i$ is the catchment utilization parameter, determined by the Health Center's \textit{Catchment PHC Utilization Rate} category (\texttt{low}, \texttt{medium}, or \texttt{high}),
  \item $U$ is the median \textit{Catchment Health Center Utilization Rate} among facilities in the same \emph{Health Center Capture Ratio} quantile as the Health Center of interest,
  \item $M$ is the number of CBHI members in the Health Center's catchment area,
  \item $B$ is the inflow parameter, applied uniformly across all Health Centers,
  \item $I$ is the historical number of visits to the Health Center by patients from outside its catchment area.
\end{itemize}

%TODO: add regression and colinearity references, add measures of colineratit (VIFs)
To estimate the values of $A_i$ and $B$, we constructed a regression equation for each facility, including its annualized historical cost, utilization metrics (\emph{Health Center Capture Ratio} quantile and $U$), catchment population ($M$), \emph{inflow} ($I$), and its catchment utilization tier. A linear regression model was then used to minimize the difference between predicted capitation payments and actual historical expenditures. The model produced three distinct values for $A$ (one for each utilization tier) and a single value for $B$ to account for \emph{inflow}. Note these correspond to the expected average cost per visit of members within the catchment area ($A$s) and from outside ($B$).

% table with the basic statistics for 2023-24
\begin{table}[!thbp]
\centering
\caption{Primary Healthcare System Overview and Utilization (July 2023 – June 2024)}
\label{tab:statistics}
\begin{tabular}{@{}l@{\hskip 4em}c@{}}
\toprule
 & \textbf{Value} \\
\midrule
\textit{Health Centers} & \textit{508} \\
\quad Non-Medicalized & 497 \\
\quad Medicalized & 11 \\
\textit{Private Health Posts} & \textit{600} \\
\quad First Generation & 552 \\
\quad Second Generation & 48 \\
\midrule
\textit{PHC Facility Density (per 1,000 CBHI members)} & \textit{1.01} \\
\quad Health Center & 0.46 \\
\quad Health Post & 0.55 \\
\midrule
\textit{Total CBHI Expenditure (in billion RWF)} & \textit{19.44} \\
\quad Health Centers & 15.84 \\
\quad Health Posts & 3.60 \\
\midrule
\textit{Total Visits (in millions)} & \textit{17.03} \\
\quad Health Centers & 13.19 \\
\quad Health Posts & 3.84 \\
\midrule
\addlinespace[0.1ex]
\midrule
 & \textbf{Median (IQR)} \\
\midrule
\textit{Cost per Visit (RWF)} & \\
\quad Health Centers & 1,199 (1,074-1,322) \\
\quad Health Posts & 883 (752-1,020) \\
\midrule
Catchment Size (CBHI members) & 20,209 (14,858-26,391) \\
\midrule
\textit{Average Monthly Visits} & \\
\quad Health Centers & 1,986 (1,555-2,608)\\
\quad Health Posts & 512 (210-742) \\
\midrule
Health Center Utilization by Catchment Members & 0.66 (0.50-0.85)\\
\midrule
Health Center Capture Ratio & 0.43 (0.33-0.53)\\
\midrule
Catchment PHC Utilization & 1.57 (1.36-1.79)\\
\midrule
Inflow Visits / Total Visits & 0.44 (0.32-0.56 \\
\bottomrule
\end{tabular}
\caption*{\small\textit{Note:} Costs associated with ambulance services and visits involving non-PHC procedures (e.g., cesarean sections) have been excluded, as these are reimbursed separately under the capitation model. Ambulance-related costs accounted for approximately 4.5\% of total spending, while non-PHC visits represented around 1\%.}
\end{table} 

\subsection{Operationalization and Adjustments}

To account for the Rwandan fiscal year, capitation parameters will be estimated each June with data from January to December of the previous year. Seasonal variations in service utilization will be considered by calculating capitation payments on a quarterly basis. While the core parameters $A_i$ and $B$ remain fixed, the other inputs to the formula, specifically utilization ($U$), \textit{inflow} ($I$), and \textit{Health Center Capture Ratio} quantile, will be adjusted using data from the same quarter of the previous year. This approach ensures that monthly payments more accurately reflect the expected temporal patterns in healthcare demand. The effect of applying or not applying these adjustments is discussed (using real data) in Section~\ref{sec:adjustment}. 

Potential differences in utilization from one year to the next must be taken into account to ensure fairness and responsiveness in payment. In cases where utilization (whether PHC utilization, capture ratio or inflow) for a given facility deviates significantly from the same quarter in the previous year (specifically, when the difference between the capitation predicted using the prior year's data and the current data exceeds $\pm$30\%), adjustments can be made to correct for under- or over-payments. These adjustments are not applied retroactively but are instead carried forward, with the difference added to or subtracted from future quarterly payments. This mechanism maintains payment stability while enabling the system to adapt to significant shifts in service delivery patterns.

\subsection{Estimated Capitation Payments}

In order to make a more efficient use of the data available to mimic what will be the operational reality, we present here the estimated capitation payments resulting on using data for the fiscal year July 2023 to June 2024 (instead of using January to December 2024 as will be used for the actual rollout). Here we discuss how the resulting capitation payment for that period would compare to the billed costs during the same period, while the next section discusses the analysis of using the estimated equation for capitation payments for the last quarter of 2024 (i.e., for a period not used for the parameter estimation, so as to mimic a situation closer to what will be the operational reality with the available data). It should be noted that the payments discussed refer to transfers made by RSSB to health facilities. However, there are additional resource allocations to these facilities originating from the government treasury.

%The capitation parameters were estimated using data from July 1, 2023, to June 30, 2024, corresponding to the full CBHI fiscal year. 
Table~\ref{tab:statistics} highlights several features of the PHC system relevant to the capitation model design and performance. Health Centers account for 77\% of all visits and 81\% of the total CBHI cost after removing ambulance costs and visits with non-PHC services, reflecting their broader service package and higher per-visit costs at those facilities. There is substantial variation in cost per visit across facilities. A notable share of visits to Health Centers are made by patients from outside the catchment area, contributing to a relatively modest median capture ratio. This high level of patient movement across catchments, combined with the variation in facility-level costs per visit, presents a key challenge in designing a fair and predictable capitation formula.

After constructing the regression dataset and fitting the linear model, the capitation parameters were estimated as shown in Table~\ref{tab:capitation}. As expected, the regression revealed strong collinearity among the predictor variables. This reflects the underlying structure of the health system, where facilities with higher utilization often also retain a larger share of visits from their catchment population. While this collinearity limits the interpretability of individual coefficients, it does not undermine the validity of the resulting payment estimates, which were the primary objective of the model. The purpose of the regression was not statistical inference, but rather to derive a set of parameters that produce capitation payments closely aligned with historical expenditure patterns.

% estimated parameters
\begin{table}[!ht]
\centering
\caption{Estimated Parameters for Capitation Formula}
\label{tab:capitation}
\begin{tabular}{@{}l@{\hskip 4em}c@{}}
\toprule
\textbf{Parameter} & \textbf{Estimated Value (RWF)} \\
\midrule
A (Low Utilization Catchments)      & 912 \\
A (Medium Utilization Catchments)   & 1,278 \\
A (High Utilization Catchments)     & 1,562 \\
\midrule
B (Inflow Component) & 1,126 \\
\bottomrule
\end{tabular}
\end{table}

Table~\ref{tab:grouping} presents the segmentation of facilities based on two dimensions: the total PHC utilization of their catchment members (which determines the \textit{$A_{i}$} parameters of the capitation formula) and the ratio of those visits captured by the Health Center. For each capture ratio group, the table shows the median utilization by catchment members at their assigned Health Center. As expected, facilities with higher capture ratios tend to have higher levels of in-center utilization by their catchment populations.

% facility groupings
\begin{table}[!ht]
\centering
\caption{Catchment and Facility Groupings}
\label{tab:grouping}
\begin{tabular}{@{}l@{\hskip 4em}c@{}}
\toprule
\textbf{Total PHC Utilization range} & \textbf{Classification} \\
\midrule
(0.79--1.44)   & Low \\
(1.44--1.70)   & Medium \\
(1.70--2.47)   & High \\
\midrule
\textbf{Capture Ratio Group} & \textbf{Median Utilization} \\
\midrule
Group 1 (0.10–0.31)     & 0.382 \\
Group 2 (0.31–0.39)     & 0.573 \\
Group 3 (0.39–0.47)     & 0.664 \\
Group 4 (0.47–0.55)     & 0.792 \\
Group 5 (0.55–0.88)     & 1.014 \\
\bottomrule
\end{tabular}
\end{table}

\begin{figure}[ht]
\centering
\includegraphics[width=1\textwidth]{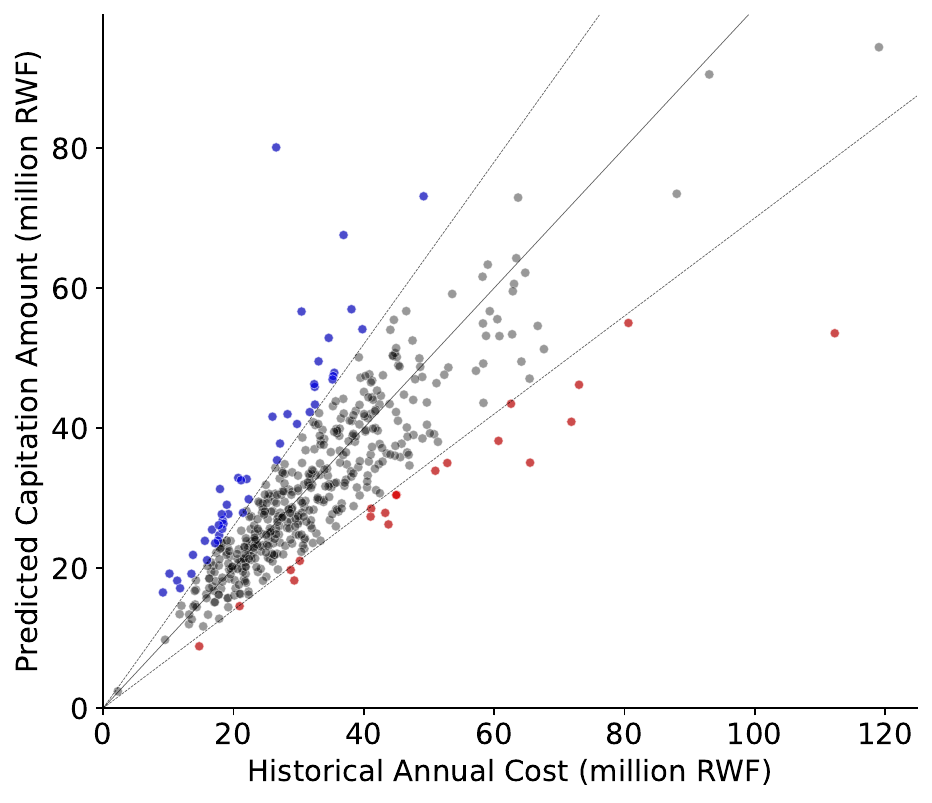}
\caption{\textit{Predicted Capitation Amount vs. Historical Costs}. The solid line indicates perfect agreement ($y = x$), while the dashed lines represent a $\pm30\%$ deviation from historical costs.}
\label{fig:capitation_performance}
\end{figure}

The performance of the estimated capitation parameters was evaluated using the July 1, 2023, to June 30, 2024, data by comparing the predicted capitation payments to the actual historical costs. Specifically, we calculated the percentage difference between the predicted and observed costs for each facility. The distribution of these differences is presented in Figure~\ref{fig:capitation_performance}, providing insight into how closely the capitation formula aligns with past expenditure patterns.

\begin{figure}[ht]
\centering
\includegraphics[width=1\textwidth]{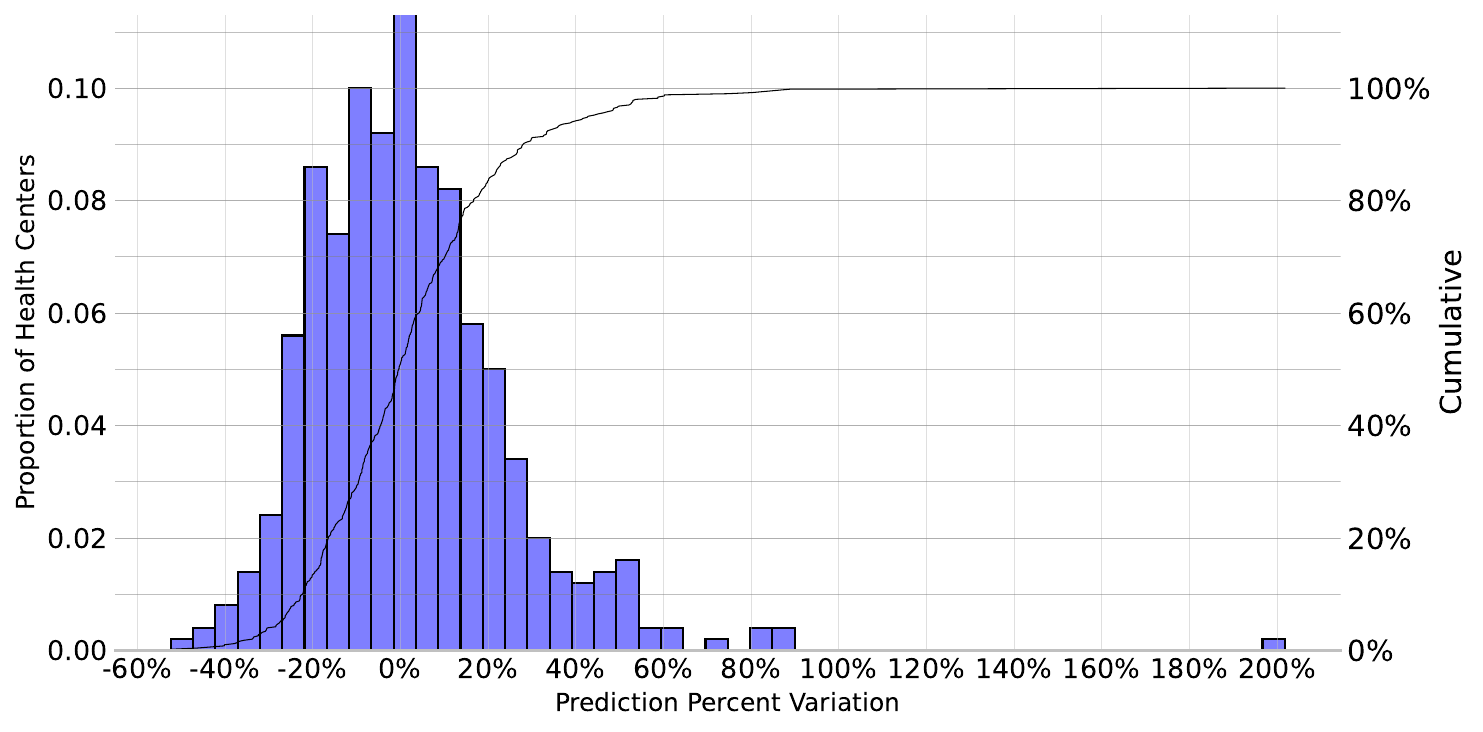}
\caption{\textit{Histogram of the Predicted Percent Variation Between the Predicted Capitation Amount vs. the Historical Costs}. The variation is computed as the difference between the predicted capitation amount and the historical cost, divided by the historical cost for each facility.}
\label{fig:percent_variation}
\end{figure}

Figure~\ref{fig:percent_variation} shows the distribution of facilities by the percent variation between their predicted capitation amount and historical cost. While the majority of facilities (around 55\%) are underpaid, the degree of underpayment is generally smaller compared to the overpayments. Notably, a larger share of facilities experience overpayments exceeding 20\% (approximately 20\%) than underpayments beyond 20\% (roughly 10\%), indicating that although the model tends to slightly underpay more facilities, the overpayments are more substantial in magnitude.

% histogram

%TODO: compare estimated parameters with average visit costs (enric) not sure I understand this one, but there's a plot of variation cost-capitation vs. cost per visit that's quite illustrative on the correlation among the two
%TODO: add time series of average costs and number of visits?

The distribution of differences between capitation payments and historical facility costs reveals several outliers, driven by structural and contextual factors. A strong correlation exists between a facility's historical cost per visit and the gap between capitation and historical reimbursements: facilities with unusually low historical costs tend to appear overpaid under capitation, while those with high costs may appear undercompensated. Among the apparent overperformers are high-volume facilities with low cost per visit, some of which attract a large share of patients from other catchment areas. A subset of these is located near hospitals and serves as de facto gatekeepers by handling referrals and triage. These facilities bear heavier workloads, but their marginal costs may remain low, which makes it difficult to capture in a standard formula (see Section~\ref{sec:alt-approaches} for further discussion on alternative approaches tested during the design phase).

On the other end of the spectrum, some medicalized health centers exhibit persistently high costs per visit, even after filtering out non-PHC services. This is likely attributable to their greater diagnostic capacity and more specialized staff, which result in higher service intensity. These variations underscore the need for RSSB to consider facility-level adjustments in selected cases to ensure the capitation model remains fair, operationally realistic, and aligned with broader health system goals.

The estimated parameters using data for January to December 2024 (as compared to the fiscal year used in this work's analysis) yield 867, 1225, and 1506 for the low, medium, and high utilization $A$ parameters and 1142 for the $B$ inflow component, i.e., all variations with respect to the 12-month period six months earlier are below 5\%, indicating the estimation is relatively stable while still showing a trend of diminishing average cost for within catchment visits while no variation to slight increment in the inflow visits.

\subsection{Adjusted Capitation Payments}\label{sec:adjustment}
%TODO: include analysis of what happens when we apply that formula to estimate capitation for first quarter of 2025. Explain how the differences can be broken into diferent utilization and inflow, expected difference and actual shifts in the parameters (average visit cost)

%This section examines the impact of implementing an adjustment mechanism in the capitation formula, aiming to align predicted payments more closely with actual healthcare costs. The capitation model was initially trained using historical data from July 2023 to June 2024, incorporating key predictors such as facility inflow and utilization patterns. However, because utilization can shift over time, particularly in facilities that experienced low historical usage but increased demand in the capitation year, the adjustment aims to correct discrepancies by incorporating real-time data. This correction is implemented quarterly, with updated information applied to subsequent payment periods.

To assess the impact of the proposed quarterly adjustments to account for utilization shifts, predictions were compared against observed cost data from August 2024 to December 2024. Since data for January to June 2025 were not yet available at the time of analysis, the corresponding months from 2024 were reused unaltered. Although drawn from the prior year, this substitution maintains consistency in seasonal patterns.

\begin{figure}[!ht]
    \centering
    \begin{subfigure}[b]{0.9\textwidth}
        \centering
        \includegraphics[width=\textwidth]{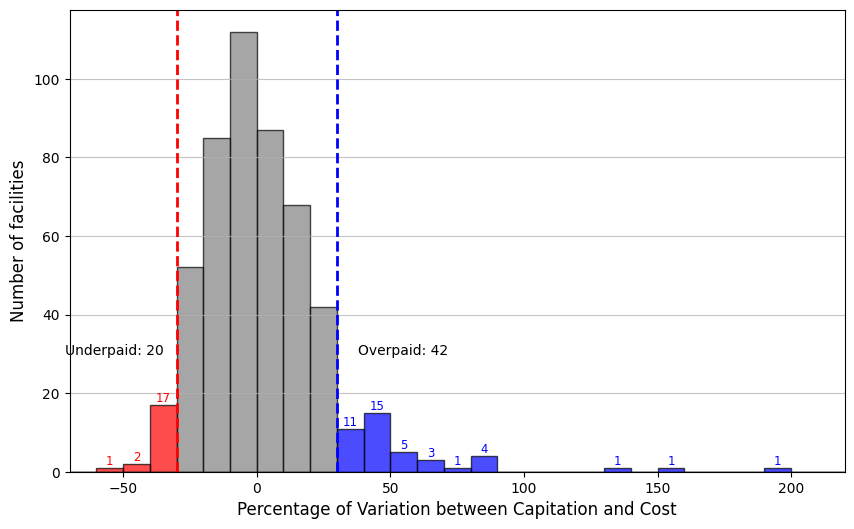}
        \caption{\textit{Without adjustment: 20 underpaid, 42 overpaid providers.}}
        \label{fig:capitation_without_adjustment}
    \end{subfigure}

    \vspace{1em} % add vertical spacing between subfigures

    \begin{subfigure}[b]{0.9\textwidth}
        \centering
        \includegraphics[width=\textwidth]{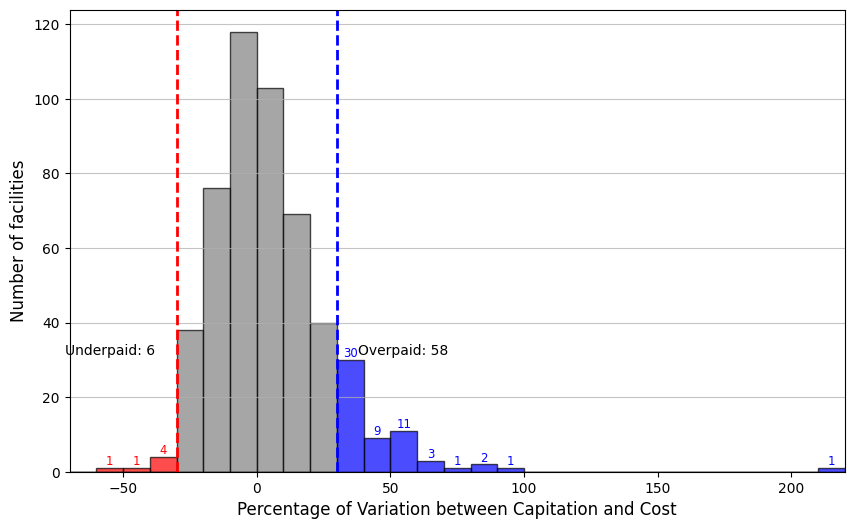}
        \caption{\textit{With adjustment: 6 underpaid, 58 overpaid providers.}}
        \label{fig:capitation_with_adjustment}
    \end{subfigure}
    
    \caption{Comparison of percentage variation between predicted capitation and actual costs, before and after applying adjustment. The adjusted model substantially decreases the number of underpaid providers with a modest increase in overpayments, indicating improved fairness and alignment with actual service costs.}
    \label{fig:capitation_variation_comparison}
\end{figure}

Facilities were classified as \textbf{underpaid} or \textbf{overpaid} based on whether the predicted capitation differed from actual costs by more than 30\%. The plots in Figure~\ref{fig:capitation_variation_comparison} illustrate the distribution of percentage deviations between predicted and actual costs under two scenarios: one without any adjustment and one with the quarterly adjustment applied. This comparison highlights the adjustment's effectiveness in reducing underpayment without excessively increasing overcompensation.

Comparing these two plots provides clear insights into the impact of adjusting capitation payments. Without adjustment, more facilities are categorized both as "underpaid" (20) or "overpaid" (42) compared to the adjusted scenario. With adjustment, the number of facilities categorized drops significantly from 20 to 6, indicating a marked improvement in fair payment distribution. The "overpaid" category increases slightly from 42 to 58 with adjustment, suggesting an acceptable trade-off to ensure fewer underpaid facilities.

%\begin{enumerate}
%    \item Without Adjustment
%        \begin{itemize}
%            \item More facilities are categorized as "underpaid" (20 providers) compared to the adjusted scenario.
%            \item "Overpaid" facilities count is lower (42) than in the adjusted scenario.   
%        \end{itemize}
%    \item With Adjustment
%        \begin{itemize}
%            \item The number of facilities categorized as "underpaid" drops significantly from 20 to 6, indicating a marked improvement in fair payment distribution.
%            \item However, the "overpaid" category increases slightly to 58, suggesting an acceptable trade-off to ensure fewer underpaid facilities.
 %       \end{itemize}
%\end{enumerate}

% Conclussion 
%While the regression parameters remain fixed (calculated from historical data), the incorporation of updated running data for capitation calculations reduces the number of underpaid providers significantly. Although it increases the count of overpaid providers, the magnitude of improvement in payment fairness justifies the adjustment approach. This strategy achieves a more balanced and equitable allocation of healthcare resources and it demonstrates the potential for improving capitation strategies by integrating real-time data.

\subsection{Robustness of the Capitation Formula and Alternative Approaches}
\label{sec:alt-approaches}

We tested the robustness of the capitation formula parameters by performing a train-test split of the dataset. The dataset was randomly divided into a training subset used to estimate the model parameters, and a test subset used to evaluate their performance. This approach allowed us to assess how well the estimated parameters generalize to unseen data. We performed 500 random train-test splits. In each iteration, model parameters were estimated on the training set and evaluated on the test set. We tracked the frequency of over- and under-payments beyond a fixed tolerance and computed the average and standard deviation of each parameter. Using the mean coefficients across splits, we applied the model to the full dataset to confirm stability and predictive consistency. Table~\ref{tab:robust_eval} shows the average value and standard deviation for the parameters estimated in the evaluation.

Several alternative approaches were explored during the design of the capitation formula, though the results are not described in this work. One involved modifying the inflow component using a binary indicator for Health Centers located within 2 km of a hospital. These facilities often act as referral gatekeepers, attracting low-cost visits from outside their catchment area. While performance improved slightly, further analysis revealed no consistent patterns in cost per visit or referral rates among these centers, so this approach was discarded due to potential implementation complexity.

A second alternative adjusted the inflow component based on the share of referred patients among inflow visits. This effectively reduced overpayment for some facilities near hospitals, but raised concerns about disincentivizing appropriate referrals, posing risks to patient outcomes under capitation. As such, it was also ruled out.

An alternative approach explored different robust regression models: Huber, RANSAC, and Theil-Sen-to estimate the capitation parameters. These methods demonstrated a more balanced distribution between overpaid and underpaid facilities compared to the linear model, which exhibited a bias toward overpayment. However, given that the primary concern during model design was to minimize the risk of underpayment, which could negatively affect service quality and facility operations, we chose to retain the linear model described in the methods section. Its moderate tendency to overestimate aligns with stakeholder preferences for a conservative approach that avoids financially disadvantaging providers.

Finally, other variables such as population density, geographic region, and staffing levels were tested but showed minimal gains in predictive accuracy and were not pursued further.

% robustness test of the formula
\begin{table}[!bhtp]
\centering
\caption{Robustness Test: Mean and Standard Deviation of Model Parameters Across 500 Splits}
\label{tab:robust_eval}
\begin{tabular}{@{}l@{\hskip 4em}c@{}}
\toprule
\textbf{Parameter} & \textbf{Mean (SD)} \\
\midrule
A (Low Utilization Catchments)  & 912 (20) \\
A (Medium Utilization Catchments) & 1,278 (21) \\
A (High Utilization Catchments)   & 1,562 (23) \\
B (Inflow Component)  & 1,126 (18) \\
\bottomrule
\end{tabular}
\end{table}

\subsection{Monitoring of Priority Indicators}
%TODO: Write this section, include all indicators (maybe in a table?), explin facility flagging (enric) revised

To ensure the capitation model delivers not only efficiency, but also safeguards the quality of care, a set of priority indicators will be closely monitored at the facility level. These indicators are designed to detect early signs of unintended consequences, such as reduced service utilization or compromised clinical practices, and allow for timely course correction.

The monthly indicators, defined by a group comprising representatives from the MoH, RSSB, and other key stakeholders, are grounded in existing health information systems, and many of them can be routinely tracked using IHBS data. These indicators reflect both service delivery volumes and clinical quality proxies, ensuring a balanced view of performance. 

While over thirty indicators have been designated to this effect, those agreed on as priority indicators that can be tracked using IHBS data include:

%TODO : is this list exhaustive? How is readmission rate defined? (enric) don't remember now if cost per visit and readmission rate were priority indicators, maybe yes. I added definitions just in case but left them as comments
\begin{itemize}
%    \item \emph{Visit cost}: Average cost per visit.
    \item \emph{Referral ratio}: Fraction of attended patients that are referred to a higher level of the health system.
    \item \emph{Admission ratio}: Fraction of attended patients that are admitted as in-patients. 
%    \item \emph{Readmission rate}: Fraction of attended patients that return to the health facility within one week.
    \item \emph{Average length of stay}: Average number of days between admission and discharge date for all in-patient admissions.
    \item \emph{Utilization by catchment members}: Average number of visits by members in the catchment divided by the total number of members in the catchment.
    \item \emph{Average number of tests per visit}. Total number of laboratory tests divided by the total number of visits.
    \item \emph{Average number of drugs prescribed per visit}: Total number of drugs prescribed divided by the total number of visits.
    \item \emph{Average number of visits with an antibiotic prescription}: Total number of visits with oral antibiotics prescribed divided by the total number of visits.
\end{itemize}

Facilities are flagged for further inspection based on statistical outlier detection across all priority indicators. Three complementary flag types are used, each designed to capture different forms of deviation from expected performance:

\begin{itemize}
    \item \textit{Self comparison:} The facility's current monthly indicator value is compared to its own historical distribution.
    \item \textit{District-level comparison:} The current monthly value is compared to the distribution of all facilities within the same district for the same month.
    \item \textit{Province-level comparison:} The facility's full historical distribution is compared to that of all facilities in the same province.
\end{itemize}

For the first two comparisons, a facility is flagged if its indicator value falls outside 1.5 times the interquartile range (IQR) from the reference distribution. The IQR, calculated as the difference between the 75th and 25th percentiles, is a robust method for detecting outliers, particularly in datasets with skewed and non-normal distributions \cite{tukey1977exploratory}.

For the third comparison, which involves full distribution similarity, we use the Bhattacharyya distance, a measure that captures both shifts in mean and changes in variance. This approach enables the detection of gradual or systemic drifts in a facility's behavior compared to provincial norms \cite{battacharyya1943measure}. 

In addition, capitation payments are retrospectively compared to what facilities would have billed under a fee-for-service (FFS) system. The differences are decomposed into three components: (1) the expected variability, consistent with fluctuations observed during the reference period used to estimate capitation amounts; (2) changes in utilization rates and patient inflows, which can inform future adjustments to payment formulas; and (3) shifts in the average cost per visit, potentially reflecting changes in service intensity or patient case mix.

This comprehensive monitoring framework enables the timely detection of unintended impacts on care quality and highlights potential deterioration in data quality. Such deterioration may occur under capitation, where incentives for accurate and timely reporting are generally weaker and more delayed than under FFS reimbursement models.

% ======================================================================
%\section{Data-Driven Insights for Behavioral Changes}
\section{Data-Driven Insights for Behavioral Changes: Antibiotic Use}
\label{sec:data-driven-insights}

One of the primary uses of the IHBS dataset is to inform data-driven strategies that enhance provider behavior and, in turn, improve the quality of care. By generating reliable, facility-level insights, the data can be used to identify patterns of overuse, underuse, or misuse of services, such as unnecessary prescriptions or over-ordering of tests, and support constructive performance dialogues with providers. This feedback loop creates opportunities to improve clinical decision-making, reduce waste, and ultimately deliver better care at a lower cost.

In particular, excessive antibiotic prescribing in primary care presents a serious public health challenge, contributing to the rise of antimicrobial resistance, unnecessary healthcare spending, and compromised patient safety. A substantial share of prescriptions are issued without clinical justification, often in cases where confirmatory diagnosis is lacking or the underlying condition is viral and self-limiting \cite{sulis,klein}. While the drivers of overprescription vary, ranging from provider habits to patient expectations \cite{kassem}, they point to a systemic issue that undermines the quality and sustainability of care. Addressing this challenge requires targeted, context-specific interventions \cite{li}. 

In Rwanda, the Intelligent Health Benefits System (IHBS) provides a unique opportunity to monitor and improve prescribing practices. By capturing detailed information on diagnoses and prescribed medications at the point of care, IHBS enables data-driven analysis of antibiotic use patterns. These insights support the development of tailored stewardship strategies aimed at promoting more rational prescribing behavior and strengthening the integrity of the primary healthcare system.

To explore potential quality issues in prescribing practices in Rwanda, we conducted an analysis of consultations for children under 15 years of age. We focused on a subset of pediatric visits in which all diagnoses within a given consultation belonged exclusively to a single diagnosis category, such as lower respiratory infections or gastrointestinal conditions (different consultations could fall into different categories). This restriction allowed for more clinically coherent comparisons across visits. The results presented in this article are restricted to visits to Health Posts in 2024; however, similar results are obtained when analyzing visits to Health Centers.

In 2024, there were 1.4 million pediatric visits in Health Posts (33.6\% of the total), representing 30.4\% of the total cost. Half of the visits (50.1\%) had only one diagnosis category, and one third (37\%) had two diagnosis categories. Overall, the most frequent diagnosis categories were upper respiratory syndromes, lower respiratory syndromes, and gastrointestinal syndromes. 

\begin{figure}[ht]
\centering
\includegraphics[width=1\textwidth]{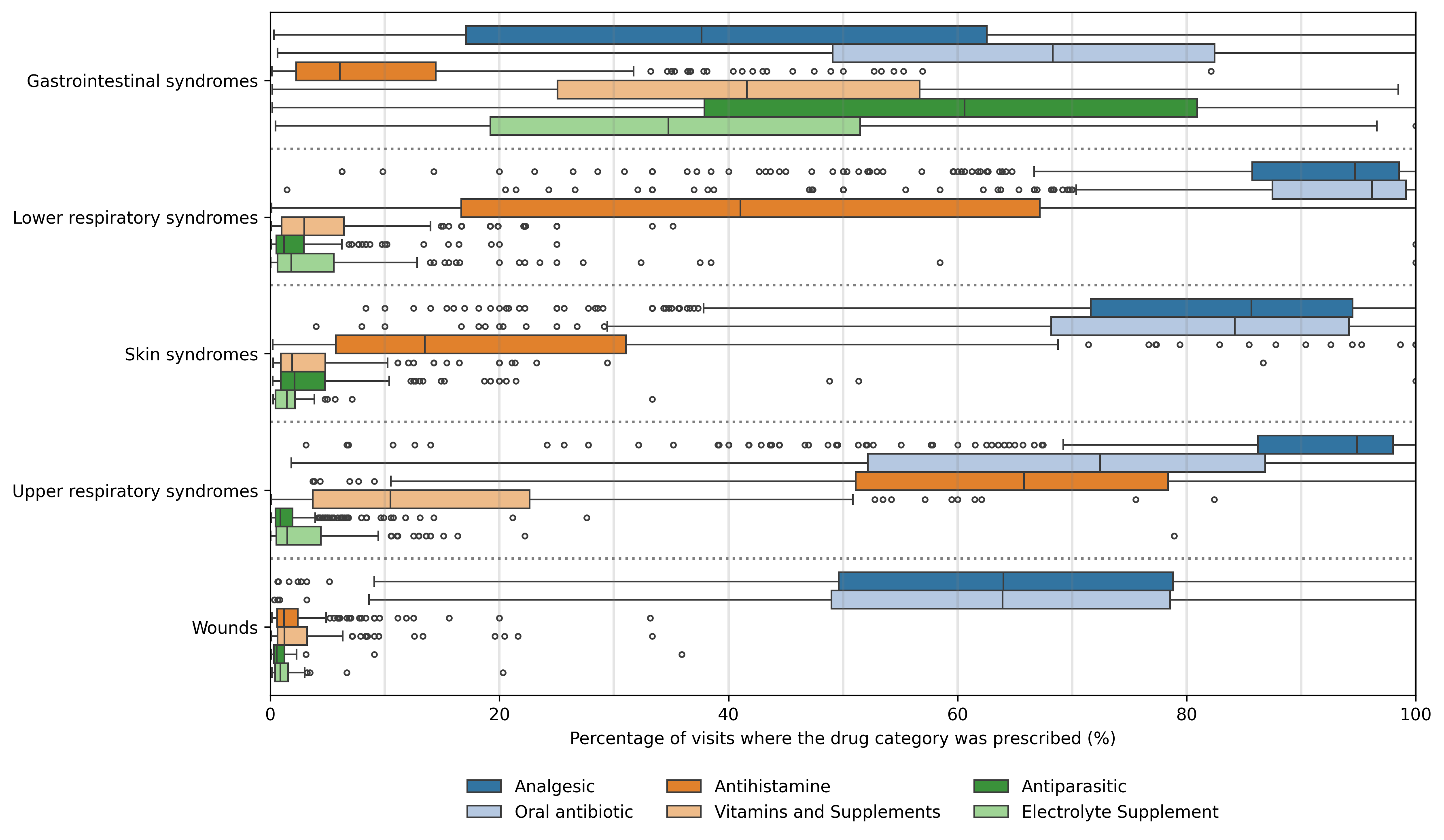}
\caption{\textit{Boxplot of health facilities on the proportion of pediatric visits in which an antibiotic was prescribed, broken by diagnosis category. Analysis was restricted to approximately 2/3 of visits with all diagnoses within one of the five categories shown in the plot.}}
\label{fig:prescription_diagnosis}
\end{figure}

Figure~\ref{fig:prescription_diagnosis} shows the distribution of facilities based on the proportion of pediatric visits, each with a single diagnosis category, in which an antibiotic was prescribed. The figure reveals widespread overuse of antibiotics, with many facilities prescribing them at consistently high rates even beyond the visible outliers. In several diagnosis categories—such as upper respiratory infections, wounds, and gastrointestinal conditions—antibiotics are prescribed in over 90\% of visits at numerous facilities, despite these syndromes often being viral or self-limiting. According to Rwanda's national pediatric treatment guidelines, such conditions typically require limited or no use of antibiotics. In addition to the high antibiotic use, the data also show widespread prescription of oral antihistamines across various syndromes, despite limited clinical justification in the pediatric guidelines. Together, these patterns suggest that a significant share of prescriptions may be unnecessary, reflecting systemic deviations from evidence-based prescribing practices.

%TODO: replace by improved figure

\begin{figure}[ht]
\centering
\includegraphics[width=0.75\textwidth]{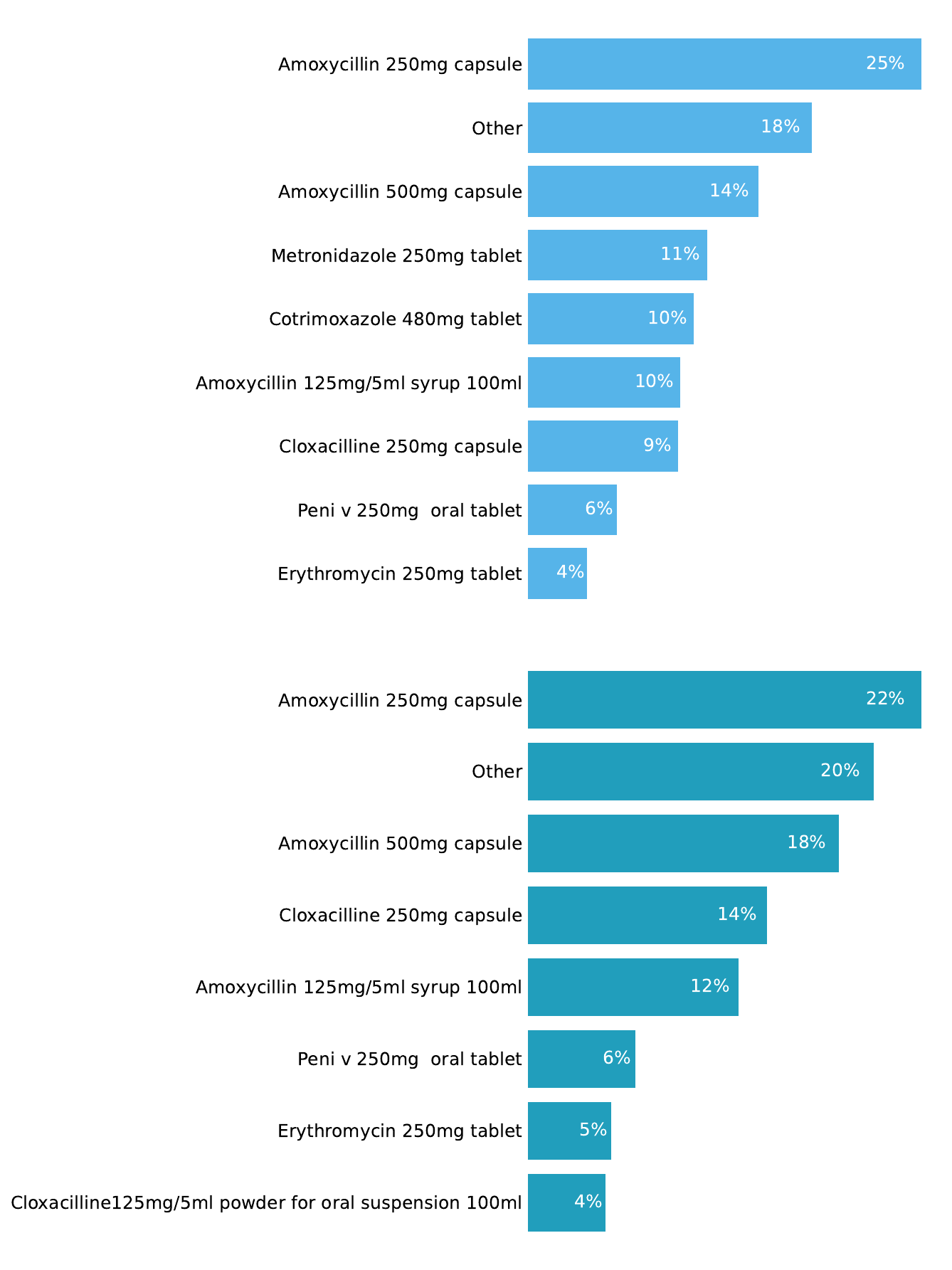}

\caption{\textit{Prescription frequency (top) and contribution to the total antibiotic cost (bottom) across all pediatric visits. The 'Other' group includes 31 different antibiotics.}}
\label{fig:antibiotics}
\end{figure}

IHBS data also highlights the opportunity to optimize the pharmaceutical supply chain by reducing the diversity of antibiotics prescribed in pediatric consultations. As shown in Figure~\ref{fig:antibiotics} (top), eight antibiotics account for over 80\% of antibiotic prescriptions. The remaining antibiotics, grouped as ‘Other’, include 31 different products. Given that these are PHC-level visits, many of these less common antibiotics likely have clinically appropriate substitutes among the most frequently used ones. Streamlining prescriptions to a smaller, standardized set could simplify procurement for Rwanda Medical Store (RMS, national procurer for public health facilities) and help improve nationwide availability. Notably, seven of the eight most frequently prescribed antibiotics also account for over 80\% of total antibiotic costs in pediatric consultations (see Figure~\ref{fig:antibiotics} bottom). Metronidazole is the only exception, due to its low unit cost.

% TODO: Stack these in different colors

\begin{figure}[ht]
\centering
\includegraphics[width=1\textwidth]{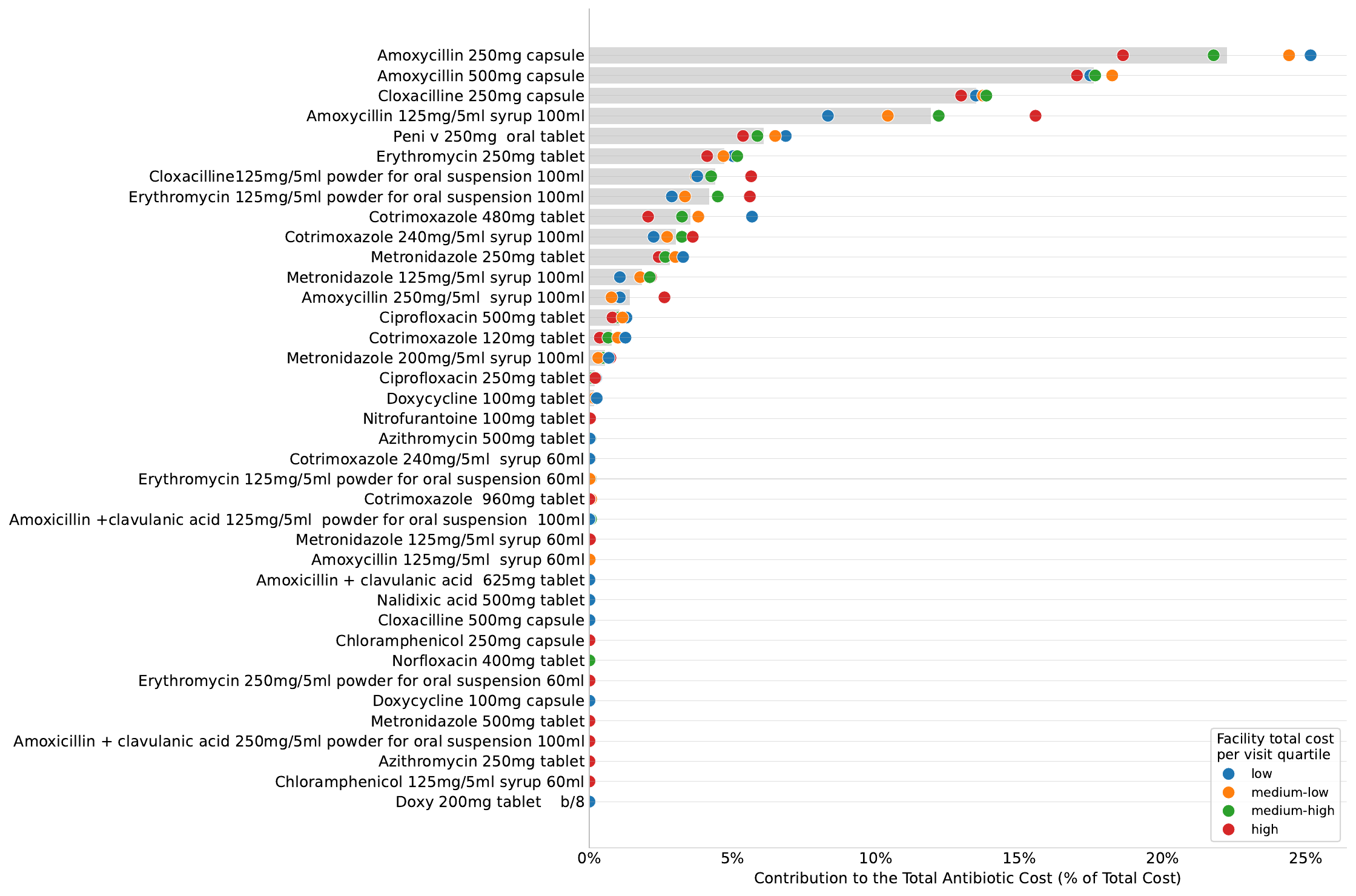}
\caption{\textit{Contribution of individual antibiotics to total antibiotic costs in pediatric consultations. The gray bars represent the average across all facilities, while the colored dots show values for facility groups segmented by overall cost per visit. This comparison highlights variations in antibiotic spending patterns across different cost strata.}}
\label{fig:antibiotic-cost}
\end{figure}

Figure~\ref{fig:antibiotic-cost} shows how the contribution of individual antibiotics to overall antibiotic costs varies across all facilities and by cost-per-visit groups. In facilities with lower overall cost per visit, the composition of antibiotic spending shifts, for example, these facilities prescribe more Amoxycillin 250mg capsules compared to the more expensive Amoxycillin 125mg/5ml syrup 100ml, which is more commonly used in higher-cost facilities. Notably, the age profile of patients is consistent across cost groups (data not shown), so these differences in prescribing cannot be explained by demographic variation. In addition to a bias toward prescribing costlier antibiotics, higher-cost facilities also tend to prescribe antibiotics more frequently and with longer courses (i.e., more pills per prescription). 

These patterns suggest that high antibiotic use and costs result from a combination of unnecessary prescriptions, preference for more expensive formulations, and longer treatment durations. This highlights a complex behavioral issue that impacts both healthcare cost and quality, which is part of a broader problem with antibiotic use globally~. Analyses like this enable us to generate facility-specific recommendations that encourage more rational antibiotic use, thereby reducing unnecessary expenditures while promoting practices that are clinically appropriate and aligned with national public health guidelines.

\section{Conclusions}
\label{sec:conclusions}
%TODO: improve this section

This article describes a methodology for the implementation and iterative refinemnt of a national-scale capitation system using high-quality health data. Rwanda's IHBS, now deployed across all public primary care facilities, enables a dynamic, data-driven financing model grounded in real-world utilization patterns. The capitation formula promotes equitable budget allocation across Health Centers and their affiliated Health Posts by accounting for intra- and inter-catchment patient flows, thereby reflecting patient mobility. Its design combines a regression-based approach with quarterly payment estimation using prior-year data to account for seasonal variations in demand. A prospective adjustment mechanism further refines fairness by correcting payments for facilities experiencing significant shifts in utilization. Early results suggest the model aligns well with historical costs for most facilities, simplifies administrative processes, and lays a foundation for quality-focused resource allocation. It is important to note that while historical reimbursement patterns are used here as a benchmark to assess the accuracy of the capitation model, they are not necessarily an ideal standard. The goal of capitation is to shift provider incentives toward efficiency and quality; some divergence from historical costs may reflect progress toward these objectives rather than a model limitation. Transparent and scalable, this framework advances Rwanda's universal health coverage goals and could offer a replicable model for other low- and middle-income countries transitioning to data-informed health financing.

Beyond payment reform, IHBS unlocks rich insights into service delivery and provider behavior, enabling evidence-based interventions to improve care quality and system efficiency. Initial analyses suggest that antibiotic overprescription is pervasive in pediatric consultations, particularly for conditions such as upper respiratory infections, for which guidelines advise against routine use. Additionally, high rates of oral antihistamine prescriptions, which are not recommended in pediatric care guidelines, further highlight misalignment with evidence-based practices. The data also reveal prescribing tendencies toward more expensive formulations, raising costs without clinical justification. Together, these patterns present opportunities to enhance stewardship, reduce waste, and align practice with national guidelines.

%IHBS also supports fraud detection and financial risk mitigation. Statistical anomaly detection models can be used to identify outlier behaviors, such as excessive visit frequency or suspicious prescriber-dispensary linkages, flagging potential integrity risks. These capabilities enable RSSB to implement responsive monitoring and feedback loops at the facility level, helping curb leakage, reinforce accountability, and maintain a balance between cost control and care quality.

As Rwanda’s digital health infrastructure continues to expand, the opportunities for data-informed system optimization will grow. The national rollout of electronic medical records (EMRs) across primary care facilities will enhance the clinical detail available in IHBS, supporting more precise quality monitoring and enabling better linkage between diagnostics, prescriptions, and outcomes. At the same time, the Rwanda Medical Supply (RMS) is deploying a new SAP-based system to manage the national health supply chain. Once integrated, this will allow for tighter coordination between procurement and actual prescribing patterns, improve inventory visibility at facility level, and further align financing with service delivery. These developments will enrich the data ecosystem underpinning Rwanda’s health financing reforms and open new avenues for quality improvement, cost control, and more equitable access to essential care.

Looking ahead, the closer integration of financing and supply chain systems presents a significant opportunity to optimize cost efficiency and ensure medicine availability. RSSB is piloting a model in which capitation payments are made net of pharmacy costs, with RMS reimbursed directly. This creates aligned incentives to reduce stockouts, increase procurement transparency, and allow RSSB to leverage IHBS data for price negotiations. Closer collaboration with RMS could also improve demand forecasting and streamline procurement by limiting the diversity of products stocked, particularly by aligning antibiotic procurement with the subset most frequently and appropriately prescribed. Such integration would enhance supply chain responsiveness and ensure equitable access to essential medicines nationwide.

In parallel, RSSB is designing a simplified reimbursement scheme for private Health Posts, offering a fixed fee per patient. As with capitation, the aim is to streamline claims processing and provide predictable revenue. However, implementation is complicated by the wide variation in costs per visit and the diversity of services offered, ranging from routine consultations to dental and ophthalmologic care. Ongoing analysis is unpacking the drivers of cost heterogeneity, including staffing, service scope, and facility scale. These insights will inform a payment structure that balances operational feasibility with fairness, ensuring that private Health Posts remain sustainable while aligning with national efficiency goals.

Ultimately, the success of the capitation reform will depend not only on payment design but also on its operationalization. Sustained monitoring using IHBS, alongside priority indicators, will be essential to flag quality deterioration or unintended changes in utilization. Equally critical will be the ability to translate analytic insights into provider behavior change. This will require continued investment in training, performance feedback systems, and aligned incentives to promote adherence to clinical standards. Continued coordination among RSSB, the MoH, and frontline providers will be essential to iteratively refine the capitation model and achieve Rwanda's vision of a resilient, equitable, and high-performing PHC system.

\section*{Acknowledgements}
%TODO: add statement of support with grant numbers

This work was supported, in whole or in part, by the Gates Foundation INV-074124. Under the grant conditions of the Foundation, a Creative Commons Attribution 4.0 Generic License has been assigned to the Author Accepted Manuscript version that might arise from this submission.

%Under the grant conditions of the Foundation, a Creative Commons Attribution 4.0 Generic License has been assigned to the Author Accepted Manuscript version that might arise from this submission. The authors wish to thank XXXXXXXXX.

%\printbibliography

% ======================================================================
%\begin{thebibliography}{99}
%\bibitem{who2023} World Health Organization: \emph{Global Monitoring Report on Financial Protection in Health} (2023)
%\bibitem{van2017capitation} vandeVen, W.P.: Risk Adjustment in European Health Care Financng. \emph{Health Policy}123(12) (2019)
%\bibitem{lazer2021data4good} Lazer, D.etal.: The Parable of Google Flu: Traps in Big Data Analysis. \emph{Science}343(6167) (2021)
%\bibitem{jia2022risk} Jia,Q., Yin,Z.: Risk Adjustment Methods for Capitation Payment: A Review. \emph{Health Services Research} 57(2) (2022)
%% --- add workshop website ---
%bibitem{sogood2025} SoGood2025 Workshop Website: \url{https://sites.google.com/view/sogood-2025/}
%\end{thebibliography}

\bibliographystyle{splncs04}
\bibliography{main} 

\end{document}